# Information-theoretic symmetry classifications of crystal patterns in the presence of noise and strong Fedorov type pseudosymmetries for an optimal subsequent crystallographic processing of these patterns

Peter Moeck

*Abstract*: Statistically sound crystallographic symmetry classifications are obtained with information theory based methods in the presence of approximately Gaussian distributed noise. A set of three synthetic patterns with strong Fedorov type pseudosymmetries and varying amounts of noise serve as examples. Contrary to traditional crystallographic symmetry classifications with an image processing program such as CRISP, the classification process does not need to be supervised by a human being and is free of any subjectively set threshold in the geometric model selection process. This enables crystallographic symmetry classification of digital images that are more or less periodic in two dimensions (2D) as recorded with sufficient structural resolution from a wide range of crystalline samples with different types of scanning probe microscopes. Alternatives to the employed objective classification methods as proposed by members of the computational symmetry community and machine learning proponents are briefly discussed in an appendix and are found to be wanting because they ignore Fedorov type pseudosymmetries completely. For sufficiently complex crystal patterns in 2D, the information theory based methods are more accurate than visual classifications by human experts.

## I. INTRODUCTION AND BACKGROUND

The symmetries of the Euclidian plane that are compatible with translation periodicity in two dimensions (2D) are tabulated exhaustively in volume A of the International Tables for Crystallography [1] and in the brief teaching edition [2] of that series of authoritative reference books from the International Union of Crystallography (IUCr). Quasicrystalline symmetries are discussed in [3]. Noncrystallographic symmetry has been defined in the on-line dictionary of the IUCr as a *"symmetry operation that is not compatible with the periodicity of a crystal pattern"* [4].

It is also noted in [4] and [5] that this term is often improperly used in biological crystallography, where one should refer either to local and partial symmetry operations on the one hand, and pseudosymmetries, on the other hand. The above mentioned on-line directory defines crystallographic pseudosymmetry simply as featuring a *"deviation"* from a space group symmetry (of one, two and three dimensions) that *"is limited"* [6], without giving an explanation on how the deviation is to be quantified. In this paper, we will provide such quantifications for three synthetic images.

Pseudosymmetry is *"a spatial arrangement that feigns a symmetry without fulfilling it"* [7] and can in direct space exist at either the site/point symmetry level of a plane symmetry group or the projected Bravais lattice type level, or a combination thereof. When a very strong translational pseudosymmetry results in metric tensor components and lattice parameters that are within experimental error bars indistinguishable from those of a higher symmetric Bravais lattice type, one speaks of a metric specialization [8].

On the site/point symmetry level, one can make a distinction between crystallographic pseudosymmetries that are either compatible with the Bravais lattice of the unit cell of the genuine symmetries or a sublattice of the genuine symmetries. These kinds of pseudosymmetries are often collectively called Fedorov type pseudosymmetries [9]. When Fedorov type pseudosymmetries exist in direct space, they will also show up in reciprocal/Fourier space. In very noisy experimental data, local and partial symmetries may become difficult to distinguish from pseudosymmetries and genuine symmetries alike.

In the presence of noise, it becomes for human classifiers of 2D crystal patterns particularly difficult to distinguish Fedorov type pseudosymmetries that are compatible with the Bravais lattice of the genuine symmetries from their genuine symmetries counterparts. The somewhat less difficult distinction of Fedorov type pseudosymmetries that are compatible with a sublattice of the underlying Bravais lattice from the genuine symmetries has been demonstrated already by this author in a very short conference paper [10]. In that case, a very simple 2D crystal pattern was used to which moderate and large amounts of noise were added. It is now time to demonstrate such distinctions for the more difficult problem in this paper.

Peter Moeck is with the Nano-Crystallography Group, Department of Physics, Portland State University, Portland, OR 97207-0751, phone: 503-725-4227; fax: 503-725-2815; e-mail: pmoeck@pdx.edu, https://orcid.org/0000-0002-5511-8482

Pseudosymmetries of the Fedorov type form plane non-disjoint from the plane symmetry group and projected Laue class of the genuine symmetries. The lowest symmetric pseudosymmetry groups will per definition always be disjoint from the lowest symmetric genuine symmetry group that provides the best fit to experimental data, but their minimal Fedorov type supergroups can be non-disjoint from that plane symmetry group. Also, per definition, the lowest symmetric plane symmetry group of the genuine symmetries, i.e. the so called "anchoring group", is the one that is measurably least broken by the generalized noise in the pattern.

Generalized noise [7] is defined in this paper as the sum of all deviations from the genuine translation periodic symmetries in a crystal's structure and/or the imaged 2D periodic properties of the crystal. At the experimental level, generalized noise as defined here combines all effects of a less-than-perfect imaging of a crystal, all rounding errors and effects of approximations in the applied image processing algorithms, effects such as uneven staining in the cryo-electron microscopy [11] of subperiodoc intrinsic membrane protein crystals, and the real structure that typically exist in addition to the ideal structure of a crystal.

Note that it is always only the ideal structure of a crystal that is strictly 2D periodic, but the symmetry group of the ideal structure is per definition the one that the real crystal possesses on average over larger sample regions which contain only a few symmetry breaking structural defects.

Fedorov type pseudosymmetry groups are by these definitions broken to a measurably larger extent than the symmetry group of the genuine symmetries (and all maximal subgroups of these symmetries). This will be further elaborated on in the second section of this paper, where a visual example is provided.

The essence of crystallographic image processing (in 2D) is the enforcing of the site/point symmetries that correspond to a correctly identified plane symmetry group on all of the pixel intensity values within the translation averaged unit cell. This is done after translation averaging by symmetrizing the structure-bearing complex Fourier coefficients of the intensity of the more or less 2D periodic image that is to be processed.

When done correctly, crystallographic image processing [12-19] increases the intrinsic[1] quality [20] of a digital image significantly by means of averaging over all asymmetric units of all unit cells of the selected image area. This enforces all site/point symmetries of a plane symmetry group onto the translation-averaged unit cell. The precondition for correctly processing an image crystallographically is the knowledge of the most likely plane symmetry that a hypothetical version of that image would possess in the absence of generalized noise. For an unknown crystal, this knowledge has historically not been easy to come by, and elucidating that kind of plane symmetry group has been a long-standing problem in both "pseudosymmetry groups", which are either disjoint or the computational symmetry subfield of computer science [21-23] and image-based electron crystallography [11-19].

The main reason that this problem had remained unsolved for more than half a century are mathematically defined inclusion relations between crystallographic symmetry groups, classes, and types [1,2]. In other words, the main reason was non-disjointness of many of the geometric models that are to be compared to the input image data and from which the best, i.e. statistically most justified, model is to be selected. Symmetry inclusion relations, non-disjointness and disjointness are explained in some detail in the third section of this paper. That section also presents the plane symmetry hierarchy tree as a visualization of disjoint and non-disjoint symmetry inclusion relationships between the translationengleiche [1,2] maximal subgroups and minimal supergroups of the plane symmetry groups.

This author presented recently a so far unique threshold-free solution to identifying the genuine plane symmetry group in a digital more or less 2D periodic image in the presence of pseudosymmetries and generalized noise [7,10,24,25]. Fedorov type pseudosymmetries do not present challenges to this solution as they are reliably identified (and can be quantified) as long as noise levels are moderate. This will be demonstrated in this paper.

The author's solution is based on Kenichi Kanatani's geometric form of information theory[2] [26-28] and complemented by analogous methods to identify the projected Laue class [7,10] of such an image as well as its projected Bravais lattice type [29]. Kanatani's theory presents a geometric "workaround" to the symmetry inclusion relations problem and has the added benefit that the prevailing noise level does not need to be estimated for the comparison of non-disjoint geometric models of digital image data. This statistical theory tackles the inclusion problem that a less restricted, e.g. lower symmetric, model of some input image data will always feature a smaller deviation (by any kind of distance measure) to the input image data than any more restricted, e.g. higher symmetric, model that is non-disjoint [26]. In other words, the fit to some experimental data with more parameters will always be better than a fit with fewer parameters.

The third section of this paper gives the relevant equations and inequalities for making objective plane symmetry and projected Laue class classifications with the author's methods so that they can be used in the fourth section. Objectivity is in this paper to be understood as only stating what digital image data actually reveal about a crystallographic symmetry without any subjective interpretation of any symmetry distance measure.

Note that information theory based crystallographic symmetry classification methods should also be developed for crystallographic symmetry classifications in three spatial dimensions because there is also subjectivity in the current practice of single crystal X-ray and neutron crystallography [7].

---

[1] The intrinsic quality as defined in [20] is the ratio of the (structural) signal to (non-structure) noise ratio with the product of the widths of the effective point spread function (spatial-angular resolution) and the square root of the number of imaging particles that contributed to the processed image. This concept is discussed further in the context of computational imaging in an appendix of this paper.

[2] According to the Merriam-Webster Dictionary, information theory is defined as *"a theory that deals statistically with information, with the measurement of its content in terms of its distinguishing essential characteristics or by the number of alternatives from which it makes a choice possible, and with the efficiency of processes of communication between humans and machines"*.
https://www.merriam-webster.com/dictionary/information%20theory

When the underlying plane symmetry in a noisy experimental image has been underestimated, i.e. only a subgroup of the most likely plane symmetry group has been identified, one does not make the most out of the available image data in the subsequent symmetry enforcing step of the crystallographic image processing procedure. On the other hand, if the plane symmetry is overestimated, "non-information" due to noise will unavoidably be averaged with genuine structural information in the subsequent processing (symmetry enforcing) of the image.

In the latter case, one may have wrongly identified a minimal supergroup of the correct plane symmetry group that the analyzed image would possess in the absence of generalized noise. That supergroup could be the union of the genuine plane symmetry group and one (or more) Fedorov type pseudosymmetry group(s).

It is, accordingly, very important to get the crystallographic symmetry classification step of the crystallographic image processing procedure just right. For that, one should only rely on the digital image data itself and refrain from any subjective considerations. The inclusion of prior knowledge into such considerations would be fine if done at a formal mathematical level.

Common practice in electron crystallography is so far, however, to make such considerations based on residuals between the translation averaged image and differently symmetrized versions of it that are based on the structure-bearing Fourier coefficients of the image intensity [11-19]. It has recently been noted with respect to cryo-electron microscopy that *"… as currently practiced, the procedure is not sufficiently standardized: a number of different variables (e.g. ... threshold value for interpretation) can substantially impact the outcome. As a result, different expert practitioners can arrive at different resolution estimates for the same level of map details."* [11].

With the author's objective and interpretation-threshold-free methods [7,10,24,25], one can now make advances with respect to the stated situation in the cryo-electron microscopy [11] subfield that deals with intrinsic membrane proteins, in the electron crystallography of inorganic materials, and the crystallographic processing of digital images in general.

It is well known that the structural (spatial-angular) resolution of crystallographic studies depends on the number of structural entities over which one averages [30]. The correct averaging can, however, only be obtained for the correct prior symmetry classification of the data that enter into such studies when no prior knowledge of the plane symmetry is available.

In crystallographic image processing on the basis of the correctly identified plane symmetry group, one enforces all of the site/point symmetries that the translation averaged unit cell image needs to feature in order to be the best representation of the input image data in the information theoretic sense. This best representation is often called the "Kullback-Leibler[3] best" or simply K-L best geometric model that the input image data maximally supports.

Compared to standard Fourier filtering [31], the processing of a digital image in the correctly determined plane symmetry group leads to a significant increase of the intrinsic image quality [20] primarily by an improvement of the structural resolution [30] of a crystallographic study. Classical Fourier filtering, on the other hand, increases the intrinsic image quality primarily by an enhancement of the signal to noise level. Note that Fourier filtering is an intrinsic part of crystallographic image processing so that the image quality is improved twice[4] by the enforcing of an objectively derived plane symmetry on an experimental image that is more or less 2D periodic.

Two different sets of structure-bearing Fourier coefficient based residuals, as implemented in the crystallographic image processing programs CRISP [13-15] and ALLSPACE [12], are most popular in the electron crystallography community. Neither of these two sets of residuals are maximal likelihood estimates combined with geometric-model selection-bias-correction terms for objective symmetry model selections of the image data. A geometric form of information theory can not, therefore, be based on these residuals in order to avoid a necessarily subjective decision of what the underlying plane symmetry most likely is (in the considered opinion of the users of these two programs).

The sets of typically employed residuals in electron crystallography quantify deviations of the experimental image with respect to differently symmetrized versions of it, but the decision which plane symmetry group is to be enforced on the input image data as part of its crystallographic image processing is with necessity left to the electron crystallographer. The CRISP program makes a suggestion that the user may either accept or overwrite, but relies heavily on visual comparisons between differently symmetrized versions of the input image data.

This author has not used ALLSPACE (in its 2dx incarnation [16]) so far as no version that runs on Microsoft Windows compatible computers seems to exist. There are also competing computer programs with less comprehensive residuals, e.g. VEC [17,18] and EDM [19] that rely even more heavily on visual comparisons of the translation averaged image to its fully symmetrized versions. (Note in passing that VEC stands for "visual computing in electron crystallography" and is aptly named with respect to its reference to a human being's vision.)

The benefits of the correct crystallographic processing of a more or less 2D periodic image will be demonstrated as a secondary goal of this paper. Scanning probe microscopists should take notice as these demonstrations are mainly directed to them. This is because crystallographic image processing is just as applicable to more or less 2D periodic images from scanning probe microscopes [25,29] as it is to images from parallel illumination transmission electron microscopes (as used in electron crystallography). Scanning probe microscopists may, however, like to correct for scanning distortions in their images of 2D periodic samples with direct space tools such as Jitterbug [32] or Smart Align [33] before they make crystallographic symmetry classifications and process their images crystallographically.

---

[3] In more technical terms, as explained in more detail below in the third section of this paper, the K-L best model features a small sum of squared residual, i.e. combined difference between the input image data and a geometric model for that data, plus twice a model-specific additive geometric-model selection-bias correction term. The K-L best geometric model within a model set for the same image input data features the minimal value of the classification specific geometric Akaike Information Criterion. The criterion itself is an estimate of the expected Kullback-Leibler divergence.

[4] For Poisson and approximately Gaussian distributed noise, the intrinsic image quality [20] increases with the enforcement of the correctly identified plane symmetry by a factor that is proportional to the square root of the multiplicity of the general position per lattice point (equal to the number of non-translational symmetry operations in the plane symmetry group). Fourier filtering alone increases the intrinsic image quality with the square root of the number of crystallographic unit cells in the recorded image for Poisson and approximately Gaussian distributed noise.

The main goal of this paper is to demonstrate the author's threshold-free classification methods on a series of three synthetic images. This may entice the computational symmetry, electron crystallography, and scanning probe microscopy communities to replace their subjectivity in crystallographic symmetry classifications with the objectivity that the information theory based methodology enables.

The limiting effects of noise and Fedorov type pseudosymmetries in more or less 2D periodic images on the accuracy of crystallographic symmetry classifications have so far rarely been analyzed. As one would expect, the distinction between genuine symmetries and pseudosymmetries of the Fedorov type becomes more difficult with increasing amounts of noise even when a geometric form of information theory is used [10]. This will be demonstrated here once more, but in much more detail, in the fourth section of this paper. That section constitutes this paper's main part and features three subsections containing nine numerical data tables as well as four figures. Two of these figures illustrate the beneficial noise reduction and crystallographic-averaging-induced structural resolution enhancement effects of crystallographic image processing.

In order to facilitate direct comparisons to results obtained by one of the two most popular traditional crystallographic symmetry classification methods of electron crystallography [13-15], *.hka files were exported from CRISP and used for the calculation of the ratios of sums of squared residuals (as parts of geometric Akaike Information Criterion (G-AIC) values) of non-disjoint geometric models for the image input data. Note that CRISP does not provide a direct estimate of the point symmetry in the amplitude map of a discrete Fourier transform, which is here referred to as the projected Laue class of a more or less 2D periodic image.

The fifth section of this paper compares the results of our three crystallographic symmetry classifications (by the author's information theory based methods) to plane symmetry group estimates by the program CRISP as applied to the same and adjacent image areas of the three synthetic images. The paper ends with a summary, conclusions, and outlook section.

The first appendix presents the hierarchy tree of the crystallographic 2D point groups that are projected (2D) Laue classes. It also gives a table on which plane symmetry groups and projected Laue classes are crystallographically compatible with each other. The second appendix expands the idea of objective projected Laue class classifications from amplitude maps of the discrete Fourier transform of a more or less 2D periodic image to electron diffraction patterns from periodic and aperiodic crystals.

The third appendix provides the equations for the calculation of ad-hoc defined confidence levels for crystallographic symmetry classifications by the author's methods into plane symmetry groups and projected Laue classes. The fourth appendix discusses crystallographic image processing in the context of computational imaging, as defined in [20]. The fifth appendix gives hypothetical scenarios/examples for the correct usage of the symmetry hierarchy trees for plane symmetry groups and projected Laue classes in objective crystallographic symmetry classifications.

The sixth appendix discusses crystallographic symmetry classifications by alternative methods. Comments on what has up to December of 2021 been achieved by others with alternative computational symmetry and machine learning approaches are provided there. That discussion extends an earlier review by this author of the whole field of direct and reciprocal/Fourier space based crystallographic symmetry classifications of digital images that are more or less 2D periodic [34].

An outlook on future developments of the information theoretic crystallographic symmetry classification and quantification methodology and their potential applications is provided in the seventh appendix. So far unpublished results of this author on the classification of parallel-illumination transmission electron microscope images from an intrinsic membrane protein in two alternative conformations are mentioned there.

The eights appendix is an account on the creation of the graphic piece of art (crystal pattern) [35] that is analyzed in this paper. It is by the artist Eva Knoll herself and clarifies that the genuine plane symmetry in Figs. 1 to 3 must be *p4* in spite of a visual *p4gm* appearance.

The latter is clearly a Fedorov type pseudosymmetry because it is measurably more severely broken than the genuine plane symmetry. The *p4* symmetry of the discussed graphic work of art is simply the result of the tiling of the Euclidian plane with digital copies of a single painted ceramic tile. This tile features the shape of a square and represents one asymmetric unit, of a square unit cell, see Fig. A-7. Four tiles represent one unit cell of the composite graphic artwork [35]. The rest of the tiling is translation periodic in 2D.

## II. FEDOROV TYPE PSEUDOSYMMETRIES ILLUSTRATED ON A NOISE-FREE SYNTHETIC IMAGE

Figure 1 shows a slightly enlarged reproduction of a 2D crystal pattern that originated with the artist Eva Knoll [35]. After expansion by periodic motif stitching of a digital representation of the original artwork as presented in [35], that pattern features in total approximately 144 primitive unit cells. The expanded image/crystal pattern is provided in the supporting material of the published version of this paper in the *.jpg format, 1160 by 1165 pixels with 24 bit depth, and 413.058 bytes at http://scripts.iucr.org/cgi-bin/paper?ou5022#suppinfoanchor. Only some primitive 16 unit cells are shown in Fig. 1.

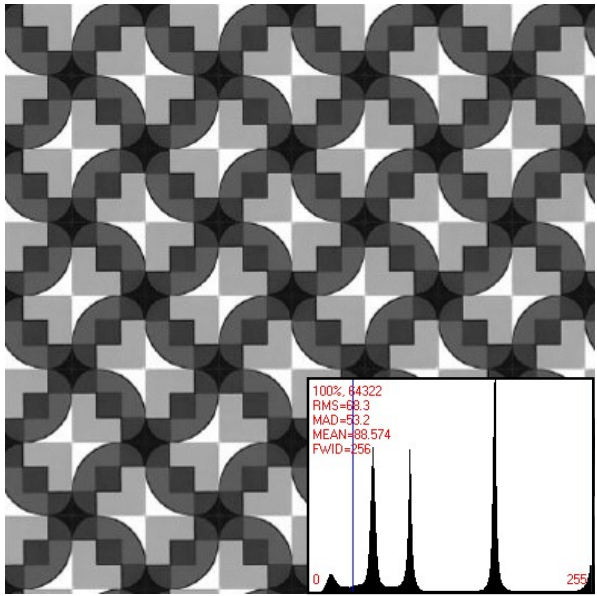

**Fig. 1.** Reproduction of a digital photo of the graphic artwork "Tiles with quasi-ellipses" (1992, acrylic on ceramic) by Eva Knoll with permission by the artist and histogram of the expanded pattern as inset. The blue thin line and descriptive annotations in red ink in the histogram are due to the program CRISP. The histogram entry "100%, 64322" means that there are 64,322 pixels with a gray level of 166, delivering the highest peak in the histogram. "RMS = 68.3" is the standard deviation (root mean square). "MAD = 53.2" is the mean absolute deviation from the mean value. "MEAN = 88.574" is the average brightness value, with 0 corresponding to black and 256 to white. "FWID = 256 is the full width of the histogram, i.e. maximal brightness value minus minimal brightness value.

The computer program Image Composite Editor [36] was used for the periodic motif stitching. (No stitching artifacts such as possible superlattice reflections were detected in the amplitude map of the discrete Fourier transform of the approximately six-fold enlarged crystal pattern that serves as the basis of this study.) There are about 15.5 periodic motifs in Knoll's paper.

The stitched/expanded digital pattern (of which Fig. 1 shows a small selection) serves in this paper as the basis of three synthetic images that are to be classified with respect to their crystallographic symmetries and Fedorov type pseudosymmetries. The two per design noisy versions of the pattern (in the series of analyzed images) are subsequently processed crystallographically in order to demonstrate the technique's benefits with respect to the noise suppression and site/point symmetry enforcing of such a processing.

Because the physical piece of graphic art from which the digital pattern in Fig. 1 was created is hand made, none of the 2D translation compatible crystallographic symmetries of the Euclidean plane are strictly speaking present (as they are only mathematical abstractions). It is, however, standard practice to assign a plane symmetry group to such a crystal pattern as one would also do for any sufficiently well resolved image from a real crystal in the real world. That symmetry group is per definition the one that is least broken by structural, sample preparation, imaging, and image processing imperfections (generalized noise). For the purpose of this demonstration, the assumption is made that the imaging and image processing imperfections of the pattern in Fig. 1 are negligible and that there are no structural imperfections/defects that are intrinsic to the imaged object. The generalized noise in that pattern is, therefore, negligible and we call this image the noise-free member of a series of three images that are to be classified with respect to their crystallographic symmetries and Fedorov type pseudosymmetries in this paper.

Note in passing that there are "very faint narrow-straight crosses" inside the "curved dark diamonds" that feature four-fold rotation points in Fig. 1. The fractional unit cell coordinates 0,0, 1,0, 01, and 1,1 are assigned to such "diamonds" in Fig. 2. The "crosses" originate from the translation periodic tiling of digital photos of the same painted ceramic square tile, Fig. A-7, without gaps or overlaps on a square lattice.

The artist might have had the intention to "amuse" (maybe also somewhat confuse) a human classifier of the art object's crystallographic symmetries by presenting a combination of genuine symmetries and strong pseudosymmetries of the Fedorov type. The approximate translation symmetry of the piece of graphic art is clearly revealed in the 2D crystal pattern in Fig. 1. Approximate point/site symmetries in the approximate unit cells can also be recognized visibly.

A human expert classifier would most likely assign plane symmetry group *p4gm* to the pattern in this image at first sight because approximate four-fold and two-fold rotation points as well as mirror and glide lines are all visibly recognizable in their required spatial arrangements in all of the more or less translation periodic unit cells of the 2D crystal pattern in Fig. 1.

The different types of visually recognizable point/site symmetries in all individual unit cells are probably broken by slightly different amounts, but these differences appear to be so minor that a human being may just assume they are all broken by exactly the same amount. Under this assumption, plane symmetry group *p4gm* would indeed underlie the completely symmetric idealization of the 2D crystal pattern in Fig. 1. The rather sharp peaks in the histogram in Fig. 1 are to be interpreted as genuine characteristics of the underlying 2D crystal pattern since no noise was added to deliberately disturb this pattern.

The image-pixel-value based classification of this pattern with the author's method reveals, however, plane symmetry groups *p2* and *p4* as genuine, with *p2* least broken being the anchoring group, and the Fedorov type pseudosymmetry groups *p1g1*, *p11g*, *c1m1*, and *c11m* as quantitatively more severely broken than the *p2* and *p4* symmetries. These pseudosymmetries combine with the genuine symmetries to the two minimal pseudo-supergroups *p2gg* and *c2mm*, as well as their respective minimal pseudo-supergroup *p4gm*.

The fourth section of this paper gives the details of the corresponding analysis. The point/site symmetry of the centers of the quite conspicuous bright "bow ties" in this pattern is visibly no higher than point symmetry group *2*, which is one of the maximal subgroups of *2mm*. Site symmetry *2mm* is, on the other hand, one of the minimal supergroups of point symmetry group *2*, but visibly more severely broken in the 2D crystal pattern in Fig. 1.

This becomes even clearer in Figs. 2 and 3. Approximately four primitive (or two centered) unit cells of this pattern in Fig. 1 are displayed in Fig. 2 after translation averaging by Fourier filtering[5].

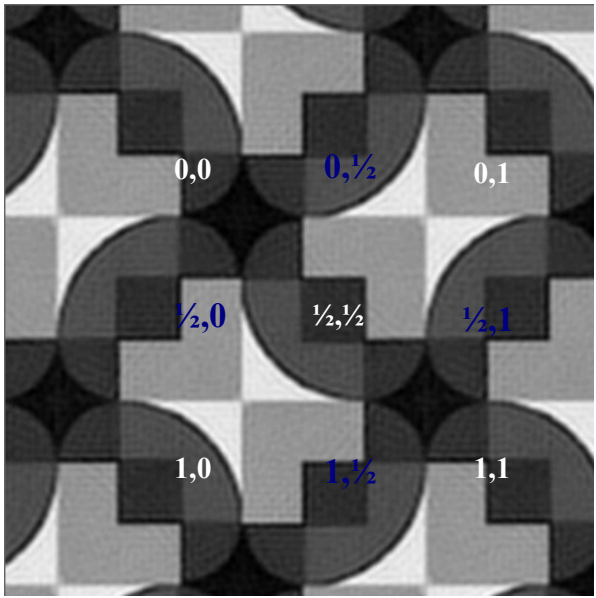

**Fig. 2.** Approximately four primitive (or two centered) translation averaged unit cells of the pattern in Fig. 1 after Fourier filtering over approximately 88 stitched-together primitive unit cells and using 956 structure-bearing Fourier coefficients for the back transform into direct space. Selected fractional unit cell coordinates as insets. Note that the "very faint narrow-straight crosses" inside the "curved dark diamonds" of Fig. 1 that are labeled here with fractional unit cell coordinates are nearly imperceptible due to a reduction of the Abbe resolution.

Note that each bright bow tie in Fig. 2 is shared between two adjacent unit cells that are based on what seems to be a square Bravais lattice. The centers of the bright bow ties are at fractional unit cell coordinates ½,0, ½,1, 0,½, and 1,½, as marked by insets in Fig. 2. These points feature visually the approximate site symmetry group *2* at best, rather than *2mm*, which would be required if the underlying plane symmetry group were to be *c2mm* or *p4gm*. The observed site symmetry *2* at these fractional unit cell coordinates is, on the other hand, compatible with plane symmetry groups *p2*, *p2gg*, and *p4*.

At the fractional unit cell coordinates 0,0, 1,0, 01, and 1,1 as well as ½,½ in Fig. 2, there are also clearly visible approximate four-fold rotation points at the centers of very dark "curved diamonds" so that a *p4* or *p4gm* classification by a human expert is probably the best anyone could come up with when the slight differences in the breaking of the individual symmetry operations are not noticed and quantified. The genuine plane symmetry group of this image can, however, only be *p2*, *p2gg*, or *p4* when the visible site/point symmetry around the centers of the bright bow ties are taken into account. (Section IV will reveal which one of the so symmetrized geometric models of the pattern in Fig. 1 is best in the minimal expected Kullback-Leibler divergence sense.)

---

[5] The obtaining of satisfactory Fourier filtering results was facilitated by the above mentioned increase in the number of unit cells in the 2D crystal pattern that underlies Fig. 1 by computational periodic motif stitching. This kind of computational increase of a digital image of the original graphic work of art in [35] is also highly beneficial to the subsequent crystallographic symmetry classification and a possible follow up step of the enforcing of the plane symmetry that most likely underlies the pattern in a statistically sound sense. Note also that Fourier filtering [31] is integral part of symmetry classifications and any subsequent crystallographic processing of a digital image. This is because the sums of squared residuals and the symmetrizing of the input image data are based only on the structure-bearing Fourier coefficients (that are periodic in reciprocal space) of a digital image.

Figure 3 zooms into the translation periodic motif of Fig. 2 and features a single bright bow tie and its immediate surrounding. Both of the arrows point in Fig. 3 to positions in the motif where the tips of the bright bow ties end and meet straight edges from the gray "right angle ruler" parts of the motif.

There is approximately a 20 % difference in the distance of these points from the horizontal and vertical edges of the gray right angle ruler shaped motif parts, so that there is definitively no mirror line from the top-right corner to the bottom left corner in this figure, as would be required for the whole motive to be part of a unit cell with plane symmetry group *p4gm*.

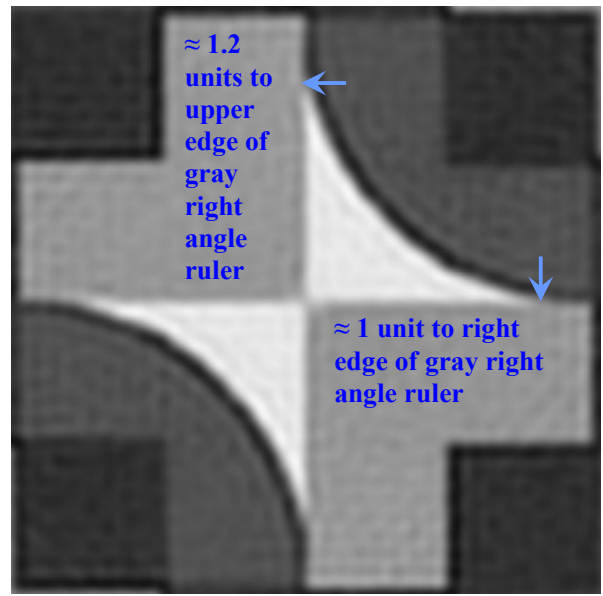

**Fig. 3.** One bright bow tie in a close up of Fig. 2. There is probably no longer an argument that the point symmetry of this feature is at best point group *2*.

Figure A-7 in the final appendix shows a photo of the original painted ceramic tile of the physical implementation of the asymmetric unit of the graphic artwork in Fig. 1. It is clear that this tile/asymmetric unit has the shape of a square. Assembling four of these tiles around a four-fold rotation point creates the unit cell of the crystal pattern in plane symmetry group *p4*. This corroborates the analysis above unambiguously. (It is irrelevant to the crystallographic analysis that the artist *intended* to create a pattern with plane symmetry group *p4gm*, see Fig. A-6 for her original sketch.)

Another example of a visual and analytical discrimination between a genuine plane symmetry group and Fedorov type pseudosymmetries in a synthetic image has been presented in [10]. The visual difference between the least broken plane symmetry and the Fedorov type pseudosymmetries in the noise-free image is clearly revealed in that short conference paper. That other paper demonstrated also that the author's methods identify the genuine symmetry and pseudosymmetry correctly in the presence of a modest amount of noise, but fail to do so for a large noise level when very small Fourier coefficients are ignored. Some of these structure-bearing Fourier coefficients were weak superlattice reflections that might easily be masked by noise in Fourier space and, therefore, overlooked. Strictly Gaussian distributed noise was used in that other paper [10].

These results were all rather similar to the results of this study as presented in the fourth section of this paper. The main differences here are that the crystallographic symmetry classifications are described in some detail, backed up by data tables, and that much more sophisticated patterns (based on a bona fide artist's work) were analyzed. As another noticeable difference to the analysis in [10], the added noise in the analyzed patterns is in this paper only approximately Gaussian distributed by design.

### III. PERTINENT EQUATIONS, INEQUALITIES, THE PLANE SYMMETRY HIERARCHY TREE; AND THEIR USAGES

As mentioned above, the structure-bearing Fourier coefficient residuals of CRISP are not in a form that allows for their usage as part of a G-AIC. Kanatani's G-AIC relies on the noise being approximately Gaussian distributed. For that kind of noise, the residuals need to be sums of squares of the differences between the input data and geometric models for that data. (Other formulations of geometric Information Criteria are possible where the generalized noise does not need to be approximately Gaussian distributed. The sums of squared residuals in Kanatani's formulation are then to be replaced by maximal likelihood estimates that are specific to that other noise distribution. The geometric-model selection-bias correction terms need to be specific to the distribution of that noise as well.)

Since crystallographic symmetry classifications are best done in Fourier space, the maximal likelihood estimates for approximately Gaussian distributed noise in more or less 2D periodic patterns takes the form of the sums of squared residuals of the complex structure-bearing Fourier coefficients for plane symmetry group classifications. For projected Laue class classifications, they take the form of the sums of squared residuals of the amplitudes of those Fourier coefficients.

The residuals are calculated as sums of squared Fourier space differences between the (Fourier) coefficients of the Fourier filtered input image data and the differently symmetrized geometric models for this data. Equation (1) gives the sum of squared residuals of the complex Fourier coefficients of a symmetrized (geometric) model of the input image data with respect to the translation-averaged-only (Fourier filtered) version of this data:

$$\widehat{J}_{cFC} = \sum_{j=1}^{N} (F_{j,trans} - F_{j,sym})^* \cdot (F_{j,trans} - F_{j,sym}) \quad (1),$$

where (.)* stands for the complex conjugate of the difference of a pair of complex numbers (.). The sum is over the differences of all $N$ structure-bearing Fourier coefficients with matching Laue indices, and the subscripts on the right hand stand for *translation averaged* and *symmetrized*, respectively. The subscript on the left hand side stands for *complex Fourier coefficients*. Note that there is a zero sum of residuals per equation (1) for the case of $F_{j,trans} = F_{j,sym}$, i.e. the translation-averaged-only model of the input image data, which features plane symmetry group *p1*.

The sum of squared residuals of the amplitudes of the Fourier coefficients is calculated in an analogous manner from the real valued amplitudes of the structure-bearing Fourier coefficients:

$$\widehat{J}_{aFC} = \sum_{j=1}^{N} (|F_{j,trans}| - |F_{j,sym}|)^2 \quad (2),$$

where the subscript on the left hand side stands for *amplitude of Fourier coefficients*.

Note again that there is a zero sum of residuals when all of the translation averaged and symmetrized Fourier coefficient amplitudes with matching Laue indices are equal to each other. This happens for the translation-averaged-only model of the input image data, which features point symmetry *2* due to the Fourier transform being centrosymmetric. Projected Laue class *2* features, accordingly, a zero sum of amplitude residuals in the data tables that are shown in the fourth section.

In order to restrict the sums of squared residuals to small numbers, the Fourier coefficients of the input image intensity and their symmetrized versions are normalized by division of the maximal amplitudes that the CRISP program provides for both the translation averaged model and the symmetrized models of the input image data in both equations (1) and (2). (That maximum is set to 10,000 for the translation averaged image data as obtained by CRISP.)

What follows below is valid for classifications into both plane symmetry groups and projected Laue classes. The same equations and inequalities as well as analogous considerations concerning the plane symmetry group hierarchy and the hierarchy of 2D point groups that are 2D Laue classes apply so that the subscripts c*FC* and a*FC* on the sums of squared residuals from equations (1) and (2) are dropped below. Two different symmetry hierarchy trees will, however, be applicable. The first one for plane symmetry groups is presented as Fig. 4 below. The second one for projected Laue classes is given in the first appendix as Fig. A1.

Kanatani's G-AIC has the general form:

$$G - AIC(S) = \widehat{J} + 2(dN + n)\widehat{\varepsilon}^2 + O(\widehat{\varepsilon}^4) + ... \quad (3),$$

where $\hat{J}$ is a sum of squared residuals, as for example given in equations (1) and (2) for the geometric model $S$, $d$ is the dimension of $S$, $N$ is the number of data points that represent the model $S$, $n$ is the number of degrees of freedom of $S$, and $\widehat{\varepsilon}^2$ is the variance of a generalized noise term, which obeys a Gaussian distribution to a sufficient approximation. The $O(\widehat{\varepsilon}^4)$ term in (3) represents unspecified terms that are second order in $\widehat{\varepsilon}^2$, while the ellipsis indicates higher-order terms that become progressively smaller.

For crystallographic symmetry classifications of more or less 2D periodic images, the dimension of the geometric models is zero (as the data are the intensity values of individual pixels that are considered to be mathematical points, i.e. zero dimensional). The degrees of freedom of the geometric models in this paper depend on the number of non-translational symmetry operations in the plane symmetry groups to which the translation-averaged input image data has been symmetrized.

They are obtained by the ratio

$$n = \frac{N}{k} \quad (4),$$

where $k$ is the above mentioned number of non-translational symmetry operations, which is equal to the multiplicity of the general position per lattice point in all plane symmetry groups.

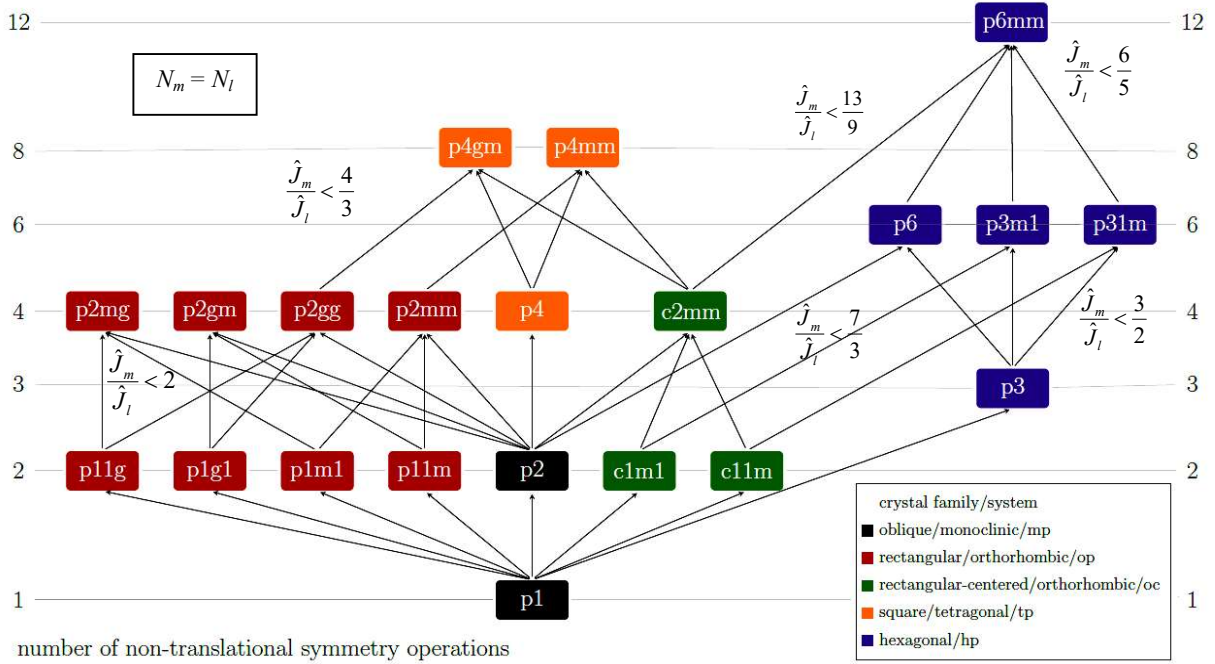

**Fig. 4.** Hierarchy tree of the translationengleiche plane symmetry groups with ratios of sums of squared complex Fourier coefficient residuals as insets. The inset ratios are valid for equal numbers of structure bearing Fourier coefficients of geometric models and apply to transitions from a $k_l$ level of the graph to a permitted $k_m$ level. Maximal subgroups are connected to their minimal supergroups by arrows. Subscript $l$ in these ratios stands for *less-symmetric* and subscript $m$ stands for *more-symmetric*.

For small and moderate amounts of generalized noise, it is justified to ignore all of the higher order terms in (3):

$$G-AIC(S) = \hat{J} + 2(dN + n)\hat{\varepsilon}^2 \quad (5).$$

The number of data points $N$ can either be constant for all geometric models or differ from model to model. The dimension of the model is defined by the type of geometric model.

Equation (5) is to be interpreted as a "balanced geometric model residual" for geometric model selections that is well suited to deal with symmetry inclusion relations. A non-disjoint and less-constrained model, which is lower symmetric, will always fit the input data better than the more constrained model that features a higher non-disjoint symmetry. The $\hat{J}$ value of the less constrained (more general) model that is in a non-disjoint relationship with a higher symmetric model will, therefore, be smaller than its counterpart for the more constrained model. In other words, the more general geometric model fits the data better than the more restricted model. This is because the more general (less constrained) model has more degrees of freedom.

The geometric model-bias correction-term $2(dN + n)\hat{\varepsilon}^2$ in (5) will for the less constrained geometric model be larger than its counterpart for the more constrained model (assuming $k > 1$ and, for convenience, an equal $N$ for both models). In other words, the better fitting, less constrained, model features a higher "geometric model selection penalty" than its worse fitting, more constrained, counterpart. This kind of interplay between fitting the input image data better at the expense of a higher model selection penalty provides the basis for objective geometric model selections by the minimizing of their G-AIC values over a complete set of geometric models.

As long as the G-AIC value of a more constrained, higher symmetric model, subscript $m$, is smaller than that of the less constrained model, subscript $l$, the former model is a better representation of the input image data than the latter:

$$G\text{-}AIC(S_m) < G\text{-}AIC \quad (6).$$

The rational geometric model selection strategy is to minimize the G-AIC values (rather than only the sums of squared residuals) for a whole set of geometric models by means of repeated applications of inequality (6). As there are two models, $S_m$ and $S_l$ in (6), one sets this inequality up for non-disjoint pairs of geometric models, one at a time, and tests if inequality (6) is fulfilled.

This allows for a more constrained/symmetric model of the input image data to be selected in a statistically sound manner as better representation of the input data although its numerical fit, as measured by its sum of squared residuals, is worse than that of the less constrained (symmetric) model. Note that the identification of which of the two geometric models is the better representation of the input image data is based solely on the input data itself and the underlying mathematics [27,28] of Kanatani's theory. (In this paper, the mathematics of crystallographic symmetries [1,2] comes obviously into play as well.)

There is no arbitrarily set threshold for the identification of the better model in the presence of a symmetry inclusion relationship, just an inequality that needs to be fulfilled numerically. All of the other crystallographic symmetry classification methods that were so far used in electron crystallography [11-19] and the computational symmetry community [21-23] feature such thresholds.

At first sight, it would seem that estimates of $\hat{\varepsilon}^2$ are needed to make objective geometric model selections by the minimization of their G-AIC values by means of inequality (6) and the definition of the first-order model selection criterion (5). Each geometric model features a different separation of the presumed geometric information amount, on the one hand, and presumed non-information (generalized noise) amount, on the other hand.

There is, however, a workaround to estimating $\hat{\varepsilon}^2$ that not only identifies the best possible separation of geometric information and non-information, but also gives an estimate of the prevailing noise in the input image data. The workarounds in this paper take advantage of the translationengleiche [1,2] symmetry inclusion relationships between plane symmetry groups as shown in Fig. 4, i.e. non-disjointness in other words.

We will explain the workaround for plane symmetry classifications in this section of the paper and refer to the first appendix for specifics that need to be considered in order to use this workaround for projected Laue class classifications on the basis of the amplitude map of the discrete Fourier transform of a more or less 2D periodic pattern.

The number of non-translational symmetry operations, $k$ is one of the two ordering principles of the hierarchy tree of the translationengleiche plane symmetry groups, Fig. 4. The other ordering principle in this figure is the non-disjointness of maximal subgroups and minimal supergroups of the plane symmetry groups. These symmetry inclusion relations are in Fig. 4 marked by arrows between maximal subgroups and minimal supergroups. The ratios of the sums of squared residuals of complex structure-bearing Fourier coefficients for climbing up from a lower level (subscript $l$ for less-symmetric) of the hierarchy to a higher level (subscript $m$ for more-symmetric) that is permitted by the special case $N_m = N_l$ of inequality (9a) are also given in Figure 4.

Translationengleich in the previous paragraph means that the addition of a non-translational symmetry operation to the unit cell of a lower symmetric group, which has the status of a maximal subgroup, results in a unit cell of a higher symmetric group, which is the former's minimal supergroup. Changes from a primitive unit cell to a centered unit cell and vice versa are permitted [37] (as they represent effectively orientation changes of symmetry operations with respect to the conventional unit cell vectors).

The translation averaged geometric model of some input image data (with plane symmetry group $p1$) is non-disjoint from the $c1m1$ symmetrized model of these data, as that plane symmetry groups is a minimal supergroup of $p1$. The centered plane symmetry group $c1m1$ with $k = 2$ is in turn in a maximal subgroup relationship with plane symmetry group $p3m1$ with $k = 6$, see Fig. 4.

The number of non-translational symmetry operations in each plane symmetry group, $k$, is given on both the left and right hand side of Fig. 4 and increases from the bottom to the top of the hierarchy tree. The connecting arrows in Fig. 4 mark plane symmetry groups as being non-disjoint. Whenever there is no arrow connecting two plane symmetry groups in Fig. 4, that pair of groups is disjoint in the translationengleiche sense [37].

In the above mentioned workaround to estimating $\hat{\varepsilon}^2$, one sets up inequality (6) for two non-disjoint models of the input image data that were symmetrized to non-disjoint plane symmetry groups, i.e.

$$\hat{J}_m + 2(dN_m + \frac{N_m}{k_m})\hat{\varepsilon}_m^2 < \hat{J}_l + 2(dN_l + \frac{N_l}{k_l})\hat{\varepsilon}_l^2 \quad (7a),$$

with $\hat{\varepsilon}_m^2 < \hat{\varepsilon}_l^2$ \quad (7b),

we obtain

$$\hat{J}_m + 2(dN_m + \frac{N_m}{k_m})\hat{\varepsilon}_l^2 < \hat{J}_l + 2(dN_l + \frac{N_l}{k_l})\hat{\varepsilon}_l^2 \quad (7c),$$

by taking advantage of both the inequality signs in (7a) and (7b).

The unbiased estimate

$$\hat{\varepsilon}_l^2 \approx \frac{\hat{J}_l}{rN_l - n_l} \quad (8),$$

for the square of the "amount" of approximately Gaussian distributed noise, where $r$ is the so-called co-dimension in Kanatani's framework (which is for our case equal to one[6]) allows us to recast inequality (7c) with $d_m = d_l = 0$ into

$$\frac{\widehat{J}_m}{\widehat{J}_l} < 1 + \frac{2(k_m - \frac{N_m}{N_l} k_l)}{k_m(k_l - 1)} \quad (9a).$$

---

[6] This is because the dimension of the data space is one (intensity values of pixels). The co-dimension is the difference between the dimension of the data space and the dimension of the model space.

The more symmetric/restricted model, subscript $m$, that fulfills inequality (7c) and (9a) will provide a better separation of noise/non-information and signal/structural information in the image, as stated in (7b).

Note that per inequality (9a), climbing up from the translation-averaged-only model of the input image data to all geometric models that have been symmetrized to minimal supergroups of $p1$ is impossible, as $k_l = 1$ in all of these cases. (There is also a zero sum of squared complex Fourier coefficient residuals for the translation-averaged-only model, equation (1), so that there is no inconsistency.)

One, therefore, simply assumes that there is indeed more than translation symmetry in the input image data and uses inequality (9a) with $k_l = 2$ and 3 as minimum. After having made that assumption, one proceeds with determining what symmetry operations there are in the input image data and to what plane symmetry group they combine.

One needs to carefully distinguish between genuine plane symmetry groups and possibly existing pseudosymmetry groups in the input image data based on the model pair's $\hat{J}_m, \hat{J}_l, k_m,$ and $k_l$ values and $N_m$ to $N_l$ ratio. Based on the definitions in the first section of the paper, the least broken symmetry at the $k_l = 2$ and 3 is the first genuine symmetry that is identified and all other genuine symmetries need necessarily be anchored on this particular symmetry group.

For the special case $N_m = N_l$, inequality (9a) reduces to

$$\frac{\widehat{J}_m}{\widehat{J}_l} < 1 + \frac{2(k_m - k_l)}{k_m(k_l - 1)} \quad (9b)[7].$$

One can then take advantage of the inequality having now the simple form of a numerical value on its right hand side that is just the sum of unity and a constant term that only depends on the difference in the hierarchy levels, $k$, of the respective two symmetrized non-disjoint models that are to be compared to each other[8], see Fig. 4. The respective ratios of sums of squared complex Fourier coefficient residuals are provided in Fig. 4 as insets for easy reference.

Inequality (9b) can be used in connection with ad hoc defined confidence levels for geometric model selections [29]. Providing such confidence levels can be understood as giving a quantitative measure of the corresponding model-selection-uncertainty, which needs to accompany any crystallographic symmetry measurement results in order to be complete [38].

The third appendix gives the formulae for calculating confidence levels for both plane symmetry and Laue class classifications. We will, however, not use inequality (9b) and equations (A-1) to (A-4) in this paper as their precondition $N_m = N_l$ is in general not met in our study. This is due to the usage of as exported *.hka files from the CRISP program or, in other words, due to the setup of this particular study. It is also possible to generalize the confidence level equations by basing them on inequality (9a).

In practice, one begins an objective crystallographic symmetry classification with calculating the sums of squared residuals for all of the geometric models that feature a multiplicity of the general position per lattice point (number of non-translational symmetry operations) of two and three, see Fig. 4. Note that plane symmetry groups $c1m1$ and $c11m$ feature two non-translational symmetry operations each, the multiplicity of the general position in the centered unit cell is four, but there are two lattice points per unit cell.

All of the geometric models with two and three non-translational plane symmetry operations are disjoint from each other per definition. Combinations of the groups with two and three non-translational plane symmetry operations lead to the majority of plane symmetry groups that are higher up in the hierarchy tree, Fig. 4.

When there is more than translation symmetry in the input image data, at least one of the geometric models that has been symmetrized to a plane symmetry group with two or three non-translational symmetry operations will have a low sum of squared residuals of the complex structure-bearing Fourier coefficients. The plane symmetry group of that model is necessarily non-disjoint from its minimal supergroups so that tests if a "climbing up" in the plane symmetry hierarchy tree is allowed by inequality (9a) can proceed until the Kullback-Leibler best geometric model of the image input data has been found. Appendix E gives a few hypothetical scenarios/examples that illustrate the logic of the climbing up procedure to identify the K-L best geometric model of some image input data for both the plane symmetry group and the projected Laue class.

By first calculating the sums of squared residuals for all eight geometric models of the input image data that feature $k = 2$ and 3, we made sure we know from which

---

The latter is zero as we are dealing with mathematical points as approximations of image pixels.

[7] Xanxi Liu and co-workers used Kanatani's G-AIC for the classification of more or less 1D periodic patterns into freeze groups in direct space. In the appendix of an earlier/longer version of X. Liu, R. T. Collins, and Y. A. Tsin, IEEE Trans. Pattern Anal. Mach. Intell. 2004, vol. 26, 354-371 (as quoted in [7] with the URL to the *.pdf), a formula analogous to (9b) (equivalent to (9a) for $N_m = N_l$) is given for direct space classifications of more or less 2D periodic patterns into plane symmetry groups. The number of analyzed translation periodic tiles enters into their equation so that a non-zero squared residual ratio exist for ascent tests of plane symmetry groups from $k_l = 1$ to $k_m = 2$ or 3, whenever more than a single tile is compared.

Working in direct space has, however, the significant disadvantage of not being able to remove much of the noise from the analysis in a simple and effective manner. Also as inconsistencies in their frieze symmetry classifications of time periodic movements of people showed [7], one has a significant pattern alignment problem in direct space that does not exist in Fourier space. In Fourier space one always works with tiles that are symmetrized to plane symmetry groups higher than $p1$ and $k_l$ is always an integer larger than unity.

[8] The comparison of two non-disjoint symmetrized models with respect of their ability to represent the input image data well is based on having an appropriate "relative measure" of their numerical distance to the common translation-averaged-only model in the first place.

plane symmetry group the anchoring and climbing up in the hierarchy tree of plane symmetry groups, Fig. 4, shall proceed as long as permitted by the fulfillment of inequality (9a).

The sums of squared residuals of the complex structure-bearing Fourier coefficients of the geometric models of the input image data that have been symmetrized to higher symmetric plane symmetry groups may be calculated on an as needed basis. Note that the whole procedure can be programmed and does not require visual inspections and comparisons of differently symmetrized versions of the input image data. This makes the information theory based classification techniques very different to the other plane symmetry classification methods that are used in contemporary electron crystallography [12-19].

Note that to conclude that a certain minimal supergroup is a plane symmetry that minimizes the G-AIC value of a geometric model of the image input data within a set of models, inequality (9a) has to be fulfilled for all maximal subgroups (and in turn their maximal subgroups). If that is not the case, that plane symmetry is only a Fedorov type pseudosymmetry as it is broken to a larger extent than the genuine plane symmetry that the hypothetical noise-free version of the input image most likely possesses. The correct crystallographic symmetry classifications of a more or less 2D periodic pattern are the plane symmetry group and Laue class that minimize the G-AIC values.

A good estimate of the amount of generalized noise can be obtained *after* the correct crystallographic symmetry classification has been made

$$\widehat{\varepsilon}_{best}^2 \approx \frac{\widehat{J}_{best}}{rN_{best} - n_{best}} \qquad (10),$$

where the subscript *best* stands for the Kullback-Leibler best model of the input image data. This estimate is in the same format as (8), i.e. the representation of the estimated noise level of the geometric model that features the lower symmetric group or class in a pair-wise model comparison procedure. When the K-L best model of the input image data has been identified, there is obviously no further climbing up allowed in the symmetry hierarchy trees of Figures 4 and A-1. This is because the G-AIC values inequality (5) can no longer be fulfilled by ways of fulfillments of inequalities (7) as well as (9a) or (9b).

Geometric Akaike weights, which are the probabilities that a certain geometric model of the input image data is indeed the K-L best model in a set of geometric models, can then be calculated on the basis of the estimated noise level and the full G-AIC values of the geometric models according to equation (5). This is not done in this paper and the reader is referred to [7,23] for details on how likelihoods of geometric models are transformed into model probabilities. Note that providing geometric Akaike weights is a second route to deriving uncertainty measures for plane symmetry group and projected Laue class classifications, without which crystallographic symmetry measurements are simply incomplete [38].

## IV. OBJECTIVE CRYSTALLOGRAPHIC SYMMETRY CLASSIFICATIONS OF A NOISE-FREE AND TWO RELATED SYNTHETIC PATTERNS WITH ADDED NOISE

As already mentioned in the introduction of this paper, crystallographic symmetry classifications are done here with both the author's methods and the electron crystallography program CRISP [13-15] using the same *.hka files of the latter program. An appropriately chosen series of these files contains all of the information on the periodic-structure bearing Fourier coefficients of the differently symmetrized geometric models of the input image data that is needed for objective classification into plane symmetry groups and projected Laue classes. In the CRISP program, these files are internally used to calculate sets of amplitude and phase angle residuals of the symmetrized input image data as well as ratios of sums of odd to even amplitudes where those are meaningful, i.e. symmetry quantifiers of the deviation of a tested plane symmetry group in reciprocal space from the Fourier filtered input image data. These files are also used internally to create symmetrized direct space versions of the input image data by Fourier back transforming for visual comparisons by the CRISP program's user.

The *.hka files can be interactively edited in CRISP. This allows for restrictions of the geometric models of the input image to a desired dynamic range of the Fourier coefficient amplitudes. The program's default value for this dynamic range is 200. (The maximal dynamic range is 10,000.)

Lowering the dynamic range leads to a reduction of the number of complex structure-bearing Fourier coefficients of the geometric models and we will make use of that for both the noise-free image and the modest amount of added noise image in the analyzed series, see Figs. 1 and 5.

Calculating the discrete Fourier transform with CRISP in its maximal dynamic range setting resulted in 3,666 complex structure-bearing Fourier coefficients for the translation averaged model of the 2D crystal pattern in Fig. 1. The patterns in Figs. 2 and 3 are, on the other hand, restricted to the back-transform of the 956 strongest complex Fourier coefficients without any symmetrizing.

A limited dynamic range of the Fourier coefficient amplitudes may lead to a reduction in the accuracy of the geometric models of the input image data. As the direct visual comparison of the 2D crystal patterns in Figs. 1 and 2 suggests, this is not a problem in the present study. Limiting the dynamic range has, on the other hand, the benefit of reducing "Fourier ripples" around features with very strong contrast changes, as can be seen in Fig. 2.

With a very large number of data points in the discrete Fourier transform of some input image data with very small amplitudes, one has to wonder if the accuracies of geometric models of the input image data are not compromised by the limited representation length of real numbers in a computer program, accumulated rounding errors, and numerical approximations in the calculation of the discrete Fourier transform.

The CRISP program allows also for restrictions of the spatial resolution of the geometric models of the input image data. This spatial resolution is defined by the Abbe criterion[9] [20].

Restricting the spatial resolution is typically necessary for noisy images that are to be classified and will be done here as well for both of the noisy images, Figs. 5 and 6. What will be called "spread noise" below is particularly effective in reducing the number of well resolved data points in a discrete Fourier transform, as demonstrated in [24], so that restrictions of the Abbe resolution need to be done in this study for the noisy images. Without judicious restrictions of the dynamical range of the structure-bearing Fourier coefficient amplitudes and the Abbe resolution of noisy images, one may produce conspicuous artifacts in the subsequent crystallographic processing of more or less 2D periodic images.

The MatLab script hkaAICnorm[10] [24], as written by a graduate student of this author, was used for the extraction of the pertinent information from the exported *.hka files. That script can be freely downloaded and calculates the sums of squared residuals, equations (1) and (2), for all of the geometric models that are used in this study from a series of *.hka files from the CRISP program.

The noise-free pattern, Fig. 1, of the synthetic image series is classified with respect to its plane symmetry group and projected Laue class in the first subsection below. The second subsection presents the classifications of the two noisy images, Figs. 5 and 6, of the series.

The results of the crystallographic processing of the two noisy patterns of the analyzed series are given in the third subsection.

*A. Classifications of the noise-free pattern in the series*

Table I lists the sums of squared residuals for a judicious selection of geometric models of the noise-free pattern, of which a small section is shown in Fig. 1. In all three analyses of this paper, circular area selections with a diameter of 1024 pixels were made in direct space for the calculation of the discrete Fourier transforms. (Circular area selections were shown to deliver better results than square selections in [24].) These sections contained approximately 88 primitive unit cells of the 2D crystal patterns that are to be classified.

**Table I:** Results of the hkaAICnorm MatLab script on the noise-free pattern that underlies Fig. 1 for geometric model selections by inequality (9a).

[9] Just like the spatial resolution limit according to Rayleigh and Sparrow, the Abbe resolution is strictly defined for an infinite imaging particle fluence [20], implying a signal to noise ratio that is independent of the spatial resolution of a system that can be modeled by a Fourier transform. It is, therefore, not fully suitable to capture the full complexity in the era of computational imaging

[10] This script works with normalized amplitudes of the structure-bearing Fourier coefficients in order to keep the numbers in the data tables below reasonably small.

| Plane symmetry group to which the image data have been symmetrized | Sums of squared residuals of complex Fourier coefficients | Sums of squared residuals of Fourier coefficient amplitudes | Number of Fourier coefficients in the geometric model of the image data |
|---|---|---|---|
| *p2* | 0.0042 | None | 956 |
| *p1m1* | 1.8799 | 0.0052 | 937 |
| *p11m* | 1.8642 | 0.0052 | 937 |
| *p1g1* | 0.0094 | 0.0052 | 934 |
| *p11g* | 0.0081 | 0.0052 | 934 |
| *c1m1* | 0.0103 | 0.0053 | 924 |
| *c11m* | 0.0110 | 0.0053 | 924 |
| *p3* | 2.5290 | 1.3339 | 954 |
| *p2gg* | 0.0096 | 0.0052 | 931 |
| *c2mm* | 0.0119 | 0.0053 | 924 |
| *p4* | 0.0065 | 0.0021 | 948 |
| *p4mm* | 1.9558 | 0.0063 | 918 |
| *p4gm* | 0.0102 | 0.0061 | 912 |

No spatial restriction was made in Fourier space for the calculation of the entries in Table I. The dynamic range of the Fourier coefficient amplitudes was set to 100 in order to restrict the number of data points $N$ in equations (1), (2), and (5) to something that is easier managed.

Note that the first seven entries in this table consist of the geometric models of the input data that feature two non-translational symmetry operations, whereas the 8[th] entry features three such operations. All of these eight models are disjoint from each other (and there are no connecting vectors between them in the plane symmetry hierarchy tree in Fig. 4).

The subsequent three entries in Table I consist of geometric models that feature four non-translational symmetry operations. The last two entries feature eight such operations and the two corresponding models are disjoint from each other (in the translationengleiche sense [37]).

The lowest sum of squared residuals of the complex Fourier coefficients is for the 2D crystal pattern that underlies Fig. 1 obtained for the geometric model that has been symmetrized to plane symmetry group *p2*. The model with plane symmetry group *p4* is listed in Table I as the one that has the lowest (non-zero) sum of residuals of the amplitudes of the Fourier coefficients.

The symmetry in the amplitude map of the discrete Fourier transform is for the *p4* symmetry model of the input image data point group *4* [1,2], which is a projected Laue class. For easy reference, the entries for geometric models with plane symmetry groups *p2* and *p4* are marked in Table I by the shading of the respective two rows.

The selection of entries in Table I has been made in order to demonstrate the climbing up from a lower level of the hierarchy of plane symmetry groups, see Fig. 4, to the next higher level.

The tests if such a climbing up is allowed by the fulfillment of inequality (9a) start always at the geometric model with the plane symmetry that has the lowest sum of squared residuals of the complex Fourier coefficients amongst the mutually disjoint models with two and three non-translational symmetry operations, i.e. the anchoring group. That starting model features always per definition a genuine symmetry, but more genuine symmetries can potentially be identified by the fulfillment of inequality (9a) for some of its non-disjoint models that may combine with the first identified genuine symmetry to some higher level genuine symmetry.

As already mentioned above, the geometric model that was symmetrized to plane symmetry group *p2* features the lowest sum of squared residual of the complex Fourier coefficients in Table I. All symmetry models that are candidates for climbing up from the geometric model that was symmetrized to *p2* in the plane symmetry group hierarchy tree, Fig. 4, i.e. *p2mm*, *p2gm*, *p2mg*, *c2mm*, and *p4* need to have a sufficiently small sum of squared residuals (and G-AIC values) with respect to all of their maximal subgroups in order to be declared genuine. Otherwise they are Fedorov type pseudosymmetries by definition. Geometric models of the input image data with low (but not the lowest) sums of squared complex Fourier coefficient residual and two or three non-translational symmetry operations may either reveal a genuine symmetry or a Fedorov type pseudosymmetry.

Plane symmetry group *p4* has only one maximal subgroup, i.e. *p2*, so that only one inequality fulfillment test is needed to find out if the former is a genuine symmetry of the 2D crystal pattern in the input image data or not. For each of the other four models mentioned in the previous paragraph, one would need to complete three inequality fulfillment tests. It is, however, already quite clear from the entries in Table I that only the models that were symmetrized to plane symmetry groups *p1g1*, *p11g*, *c1m1*, and *c11m*, have reasonably low sums of squared residuals (and G-AIC values) to make them reasonable candidates for climbing up tests. The geometric models that were symmetrized to *p2mm*, *p2gm*, and *p2mg* are just too far from fulfilling inequality (9a) with respect to the anchoring group as their sums of squared residuals in Table 1 are very much larger than that for the *p2* model.

Table II gives the ratios of the sums of squared residuals of the complex Fourier coefficients for the non-disjoint models of Table I (left hand side of inequality (9a) in the second column) together with the maximal value that these ratios may have (right hand side of inequality (9a) in the third column) in the context of minimization of the G-AIC value of the higher symmetric model of a pair of non-disjoint geometric models of the input image data. The tests if climbing up to the next level of the plane symmetry hierarchy tree is allowed consist in a simple comparison of the numerical values in the second and third column of Table II, which is recorded in the fourth column.

There is only one unconditional "yes" in the fourth column of this table, as marked by the shading of the corresponding row, so that the conclusion has to be drawn that the geometric model which has been symmetrized to plane symmetry group *p4* features the only other genuine symmetry in the 2D crystal pattern that underlies Fig. 1, i.e. the noise-free image of the series.

**Table II:** Numerical values of ratios of sums of squared residuals of the complex Fourier coefficients of non-disjoint models of the noise-free pattern that are either within their maximal allowance or not.

|  | Left hand side of (9a) | Right hand side of (9a) | Inequality (9a) fulfilled? |
|---|---|---|---|
| *p2gg* over *p2* | 2.285714 | 2.0261506 | no, blocking ascent |
| *p2gg* over *p1g1* | 1.021277 | 2.0032312 | yes, but due to pseudosymmetry |
| *p2gg* over *p11g* | 1.185185 | 2.0032312 | yes, but due to pseudosymmetry |
| *c2mm* over *p2* | 2.83333 | 2.0 | no, blocking ascent |
| *c2mm* over *c1m1* | 1.155340 | 2.0 | yes, but due to pseudosymmetry |
| *c2mm* over *c11m* | 1.081818 | 2.0 | yes, but due to pseudosymmetry |
| *p4* over *p2* | 1.547619 | 2.008368 | yes |
| *p4mm* over *p4* | 300.8923 | 1.3438819 | no, blocking ascent |
| *p4gm* over *p4* | 1.569231 | 1.3459916 | no, blocking ascent |
| *p4gm* over *p2gg* | 1.06250 | 1.3401361 | yes, but due to pseudosymmetry |
| *p4gm* over *c2mm* | 0.857143 | 1.3376623 | yes, but due to pseudosymmetry |

It is important to realize that all genuine symmetries above the $k = 2$ and 3 level must by definition be anchored on the least broken plane symmetry group, i.e. the one with the lowest sum of squared residuals for the complex Fourier coefficients at the $k_l = 2$ and 3 levels in Fig. 4. The fulfillment of inequality (9a) for a pair of non-disjoint geometric models that does not fulfill this overwriting requirement can per definition only signify a strong Fedorov type pseudosymmetry, which will be clearly visible in a direct space image. Plane symmetry groups *p2gg* and *c2mm* must be Fedorov type pseudosymmetries of the pattern in Fig. 1 because climbing up from *p2* is not permitted, see first and fourth entry in Table II. Their maximal subgroups *p1g1*, *p11g*, *c1m1*, and *c11m* are also Fedorov type pseudosymmetries as they are disjoint from the *p2* anchoring group.

Note that climbing up tests for strong Fedorov type pseudosymmetries to the $k_m = 4$ level, i.e. *p2gg* and *c2mm*, and up to $k_m = 8$, i.e. *p4gm*, have rather low values for the left hand side of inequality (9a) in Table II. This is due to the corresponding sums of squared complex Fourier coefficient residuals for the matching $k_l = 2$ and 4 levels being on the same order of magnitude in Table I. The ratios of such sums may for strong Fedorov pseudosym-metries even fall below unity, as shown for the last entry in Table II.

Ideally, one would base all calculations on symmetrized models of the input image data that feature exactly the same appropriately indexed Fourier coefficients and number of such coefficients. To obtain the same number of data points (complex Fourier coefficients) in all geometric models of the input image data, one would need to treat Fourier coefficients that are absent in certain geometric models as featuring zero amplitude and arbitrary phase. The absences can either be systematic or incidental. In both cases the zero amplitude Fourier coefficients are characteristics of the properly symmetrized geometric models of the input image data. The smallest possible entry in the second column of Table II should for genuine symmetries then be restricted to unity (in the absence of generalized noise including small calculation errors) and one can also give confidence levels for the classification into genuine symmetries by using equations (A-1) to (A-4) and provide a complete [38] crystallographic symmetry measurement result.

Relying on the *.hka files of CRISP in this study without further editing, this is, however, almost impossible to achieve in practice. The geometric models that are represented by *.hka files with different numbers of data points, different dynamic ranges, and Abbe resolutions [20] do not necessarily give always the best possible symmetrized version of the input image data in Fourier space. For the purpose of the demonstrations in this paper and to allow for the comparison of classification results that were obtained with the information theory based methods to those of the CRISP program, the accuracy of all geometric models is deemed to be more than sufficient.

On all accounts, the geometric models that CRISP provides in the form of exportable *.hka files are always quite representative of symmetrized versions of analyzed images as demonstrated by the successes of numerous electron crystallography studies [13-15] in spite of necessarily different choices for the dynamic range, Abbe resolutions [20], and numbers of included Fourier coefficients.

The identification of the projected Laue class that minimizes the G-AIC value for the 2D crystal pattern that underlies Fig. 1 proceeds analogously. Laue class *4* has already been identified above as the point symmetry of the amplitude map of the geometric model that has been symmetrized to plane symmetry group *p4*. Because the *p4* model has the lowest squared Fourier coefficient amplitude residual in Table 1, point group *4* is the anchoring point group for the Laue class classification of the pattern in Figure 1. Both this projected Laue class and 2D Laue class *2mm* feature four point symmetry operations, $k_l = 4$, and are disjoint from each other, see hierarchy tree in Fig. A-1 in the first appendix.

Table III gives the ratios of the sums of the squared Fourier coefficient amplitude residuals for the non-disjoint models of Table I (with $k_l = 4$) together with the maximal value that these ratios may have for a climbing up to the $k_m = 8$ level. Obviously, one cannot climb up from Laue class *4* to the non-disjoint Laue class *4mm* with $k_m = 8$ (in Fig. A-1), based on the numbers in this table.

Based on the low sums of squared Fourier coefficient amplitude residuals in Table I, the models for projected Laue classes *2mm* and *4mm* reveal pseudosymmetries in the input image data. This is fully consistent with the identified Fedorov type pseudosymmetries at the plane symmetry group level.

To conclude this subsection: plane symmetry group *p4* (which contains *p2* as its only maximal subgroup) and Laue class *4* are identified as both genuine in the 2D crystal pattern and crystallographically consistent with each other. The identified Fedorov type pseudosymmetries at the lowest level of the hierarchy tree of plane symmetry groups are *p1g1*, *p11g*, *c1m1*, and *c11m*. These pseudosymmetries combine with each other and the identified genuine symmetries to the pseudosymmetry groups *p2gg, c2mm*, and *p4gm*. There are corresponding *2mm* and *4mm* pseudosymmetries in the amplitude map of the noise free pattern in Fig. 1, but no *4mm* pseudo-site symmetry in the direct space unit cell of the input image data since the *p1m1* and *p11m* models of this data feature sums of squared complex Fourier coefficient residuals that are way too large to pass climbing up tests in the plane symmetry hierarchy tree of Fig. 4.

**Table III:** Numerical values of ratio of the sums of squared amplitude residuals of non-disjoint geometric models of the noise free pattern that are either within their maximal allowance or not.

| | Left hand side of inequality (9a) | Right hand side of inequality (9a) | Inequality fulfilled? |
|---|---|---|---|
| *4mm* over *4* (in c2mm setting) | 3 | 1.3438819 | no, as it should |
| *4mm* over *4* (in p2gg setting) | 2.90476 | 1.3577236 | no, as it should |
| *4mm* over *2mm* (in p2gg setting) | 1.2115385 | 1.3379878 | yes, but due to pseudo-symmetry |
| *4mm* over *2mm* (in c2mm setting) | 1.1886792 | 1.3246592 | yes, but due to pseudo-symmetry |

*B. Classifications of the two noisy patterns of the series*

Figures 5 and 6 show sections of the two synthetic patterns that were obtained by adding approximately Gaussian distributed noise to the synthetic pattern that served as basis of Fig. 1, i.e. the approximately 144 periodic motif repeats containing expanded representation of the original graphic artwork [34] that is considered to be free of generalized noise. The freeware program GIMP [39] was used to add the noise.

Spread noise swaps individual pixel intensities in the horizontal and vertical directions by a selected number of pixels. Strictly Gaussian distributed noise only changes the individual pixel values but not their positions in the translation periodic unit cell. The employed mixtures of strictly Gaussian distributed noise and spread noise add up to approximately Gaussian distributed noise. The strictly Gaussian distributed noise had been added to the 2D crystal pattern in Fig. 1 before the spread noise was added with GIMP.

The effects of the added noise are clearly visible in Figs. 5 and 6 and their histogram insets when compared to the histogram inset in Fig. 1 and that figure itself. Compared to Fig. 5, there is approximately five times as much added noise in Fig. 6.

This approximate 1 to 5 ratio of the amounts of noise is similar to that of the two noisy synthetic images in [10]. One should, therefore, expect similar crystallographic classification results as in [10], i.e. the correct distinction between genuine symmetries and Fedorov type pseudosymmetries for the image with the moderate amount of noise and a misidentification for the image with the large amount of noise.

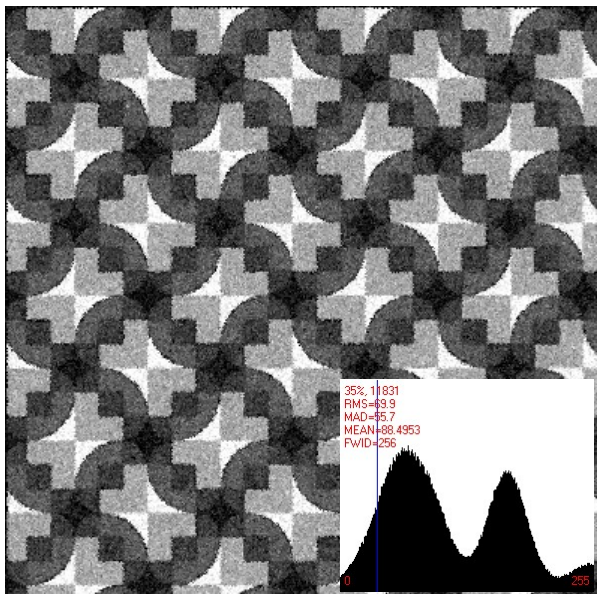

**Fig. 5.** Section of the underlying 2D crystal pattern of Fig. 1 with a moderate amount of approximately Gaussian distributed noise added. The histogram of the whole pattern is provided as inset. The blue thin line and descriptive annotations in red ink in the histogram are provided by the CRISP program. The histogram entry "35%, 11831" means that 11,831 pixels deliver the highest peak in this histogram.

We classify the noisy 2D crystal pattern in Fig. 5 first. The dynamic range in the employed *.hka file from CRISP was set to 100. The selection in Fourier space was set to a 350 pixel radius (out of the maximal possible 512 pixels), which results in a Abbe resolution reduction, that may be appreciated by a comparison of the crystallographically processed version of this image, Figure 7, in the third subsection below to the 2D crystal pattern in Fig. 1. Both of these settings resulted in a reasonable number of Fourier coefficients in the last column of Table IV.

The geometric model with plane symmetry group *p2* features again the lowest sum of squared residuals of the complex Fourier coefficients in this table. Also as before, the model that was symmetrized to plane symmetry group *p4* features the lowest sum of Fourier coefficient amplitude residuals. Again, the rows for these two geometric models of the input image data are highlighted in Table IV for easy reference by shading.

Analogous to Table II, Table V gives the ratios of the sums of the squared residuals of the complex Fourier coefficients for climbing up tests.

There are four unconditional "yes" in Table V when the prior information on the objective symmetry classification of the noise-free pattern of the crystal pattern series from the previous subsection is not used. The rows of the corresponding entries are again marked by the shading.

The preliminary conclusion from the shaded rows in Table V is that the genuine plane symmetry group of the noisy crystal pattern in Figure 5 must either be p2gg or p4. Both plane symmetry groups are disjoint from each other, see Figure 4a, so that one of these two groups has to be a Fedorov type pseudosymmetry per definition. The decision about which of these two plane symmetries is genuine relies on the necessity of the consistency of the plane symmetry classification with the Laue class classification of the noisy pattern in Figure 5.

**Table IV:** Results of the hkaAICnorm MatLab script on the modest amount of noise added image that underlies Fig. 5 for geometric model selection by G-AIC value ratios.

| Plane symmetry group to which the image data have been symmetrized | Sums of squared residuals of complex Fourier coefficients | Sums of squared residuals of Fourier coefficient amplitudes | Number of Fourier coefficients in the geometric model of the image data |
|---|---|---|---|
| *p2* | 0.0041 | None | 665 |
| *p1m1* | 1.7207 | 0.0041 | 654 |
| *p11m* | 1.7210 | 0.0041 | 654 |
| *p1g1* | 0.0059 | 0.0041 | 652 |
| *p11g* | 0.0066 | 0.0041 | 652 |
| *c1m1* | 0.0081 | 0.0043 | 655 |
| *c11m* | 0.0081 | 0.0043 | 655 |
| *p3* | 2.0554 | 1.3052 | 685 |
| *p2gg* | 0.0066 | 0.0041 | 650 |
| *c2mm* | 0.0102 | 0.0043 | 655 |
| *p4* | 0.0040 | 0.0015 | 648 |
| *p4mm* | 1.7934 | 0.0050 | 644 |
| *p4gm* | 0.0074 | 0.0050 | 640 |

**Table V:** Numerical values of ratios of sums of squared residuals of the complex Fourier coefficients of non-disjoint models of the pattern with a moderate amount of added noise that are either within their maximal allowance or not.

| | Left hand side of (9a) | Right hand side of (9a) | Inequality fulfilled? |
|---|---|---|---|
| *p2gg* over *p2* | 1.6097561 | 2.0225564 | yes, but due to pseudosymmetry |
| *p2gg* over *p1g1* | 1.1186441 | 2.0030675 | yes, but due to pseudosymmetry |
| *p2gg* over *p11g* | 1.0 | 2.0030675 | yes, but due to pseudosymmetry |
| *c2mm* over *p2* | 2.4878049 | 2.0 | no, blocking ascent |
| *c2mm* over *c1m1* | 1.2592593 | 2.0 | yes, but due to pseudosymmetry |
| *c2mm* over *c11m* | 1.2592593 | 2.0 | yes, but due to pseudosymmetry |
| *p4* over *p2* | 0.9756098 | 2.0255639 | yes |
| *p4mm* over *p4* | 448.35 | 1.3353909 | no, blocking ascent |
| *p4gm* over *p4* | 1.85 | 1.3374486 | no, blocking ascent |
| *p4gm* over *p2gg* | 1.1212121 | 1.3384615 | yes, but due to pseudosymmetry |
| *p4gm* over *c2mm* | 0.7254902 | 1.3409669 | yes, but due to pseudosymmetry |

The anchoring Laue class is point symmetry group *4* because the corresponding *p4* symmetrized model of the noisy pattern in Figure 5 features in Table IV the lowest sum of squared residuals of the Fourier coefficient amplitudes. The point symmetry in the amplitude maps of the discrete Fourier transforms of the geometric models of the crystal pattern in Fig. 5 that were symmetrized to plane symmetry groups *p1m1*, *p11m*, *p1g1*, *p11g*, *c1m1*, *c11m*, *p2gg*, and *c2mm* is point symmetry/Laue class *2mm* (Aroyo, 2016, Hahn, 2010).

Table VI is analogous to Table III and lists the ratios of sums of squared Fourier coefficient amplitude residuals for the modest amount of added noise pattern in Figure 5. The conclusion from this table is that projected Laue class *4* is the only genuine class as climbing up from the anchoring class to Laue class/point group *4mm* is not allowed.

Note that point symmetry group *4* captures the symmetry in the amplitude map of the discrete Fourier transform of the noisy crystal pattern that underlies Figure 5 better by more than a factor of 2.7 than point group *2mm*, which is at the same $k_l = 4$ level of the hierarchy tree of Figure 4b. It is, therefore, without any doubt the point symmetry of the Kullback-Leibler best geometric model of the amplitude map of that pattern.

Laue class *2mm* is accordingly to Table VI a pseudo-symmetry at the point symmetry level and the correspond-ding plane symmetry group *p2gg* can also only be a strong Fedorov type pseudosymmetry. With point group *2mm* identified as pseudosymmetry and point group *4* a genuine symmetry in the amplitude map of the discrete Fourier transform of the noisy pattern in Figure 5, there must also be a *4mm* pseudosymmetry in this map. This is confirmed by the numerical values in Table VI.

**Table VI:** Numerical values of ratio of the sums of squared amplitude residuals of non-disjoint models of the moderate amount of added noise pattern that are either within their maximal allowance or not.

| | Left hand side of inequality (9a) | Right hand side of inequality (9a) | Inequality fulfilled? |
|---|---|---|---|
| *4mm* over *4* (in c2mm setting) | 3.333333 | 1.3353909 | no, as it should |
| *4mm* over *4* (in p2gg setting) | 3.333333 | 1.3374486 | no, as it should |
| *4mm* over *2mm* (in p2gg setting) | 1.2195122 | 1.3384615 | yes, but due to pseudo-symmetry |
| *4mm* over *2mm* (in c2mm setting) | 1.1627907 | 1.3389313 | yes, but due to pseudo-symmetry |

Note in passing that the ratio of the sums of squared residuals of the complex Fourier coefficients is for the "*p4* over *p2*" row of Table V smaller than unity. This is probably the result of both small accumulated calculation errors in the analysis and slight differences in the accuracy of the representation of the geometric models in the employed *.hka files from CRISP.

There is, again no *4mm* pseudo-site symmetry in the direct space unit cell of that crystal pattern because ascent from the geometric model that was symmetrized to plane symmetry group *p4* to its counterpart with plane symmetry *p4mm* is blocked in Table V by a very wide margin.

Clear distinctions between genuine symmetries and Fedorov type pseudosymmetries were, thus, again obtained. The added approximately Gaussian distributed noise presented no challenge to the crystal pattern classification task with respect to its crystallographic symmetries when the amount of noise was modest.

The preliminary issue which of the two disjoint plane symmetry groups, *p2gg* or *p4*, is the symmetry of the Kullback-Leibler best model of the noisy pattern in Figure 5 was straightforwardly resolved by recognizing point symmetry *4* as the anchoring Laue class. Note that no prior knowledge of the classification of the noise-free pattern in the series of crystal pattern from the previous subsection was used to reach the final conclusions. As expected, the effect of adding noise is an obscuring of the differences in the amounts of breakings of the various plane symmetry groups. Adding larger amounts of noise that is to a lesser approximation Gaussian distributed should confirm the general trend that genuine symmetries and pseudosymmetries in crystal patterns get more difficult to distinguish. As we will see below, this is indeed the case.

The dynamic range of the amplitudes of the Fourier coefficients of the 2D crystal pattern in Fig. 6 was for the analysis increased to 200 and the radius of the circular area in Fourier space restricted to 200 pixels because several Fourier coefficients were essentially "washed out" by the large spread noise component of the added noise. These settings ensured that the geometric models of this pattern are represented in Fourier space by a reasonable number of data points in the fourth column of Table VII.

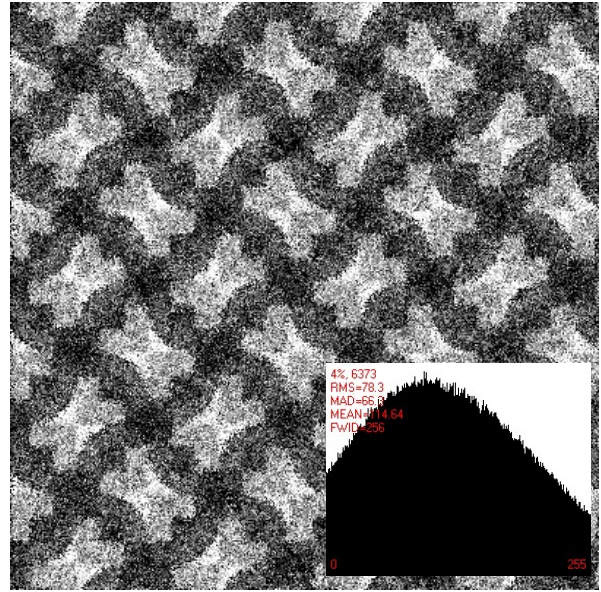

**Fig. 6.** Section of the underlying 2D crystal pattern of Fig. 1 with a large amount of approximately Gaussian distributed noise added. The histogram of the whole pattern is provided as inset. The descriptive annotations in red ink in the histogram are provided by the CRISP program.

In analogy to Tables I and IV, Table VII gives the characteristics of the geometric models for the rather noisy 2D crystal pattern that underlies Fig. 6. The two rows for the plane symmetry and projected Laue class anchoring groups are again highlighted in this table by shading of the respective rows.

Table VIII reveals, however, that genuine symmetries at the plane symmetry group level can no longer be distinguished from Fedorov type pseudosymmetries as the result of the large amount of added noise. All rows with an unconditional 'yes' in this table are, therefore, shaded.

Plane symmetry group *p4gm* is now misidentified as genuine and the symmetry that most likely underlies the rather noisy 2D crystal pattern in Fig. 6. Note that climbing ups in the plane symmetry hierarchy tree of Fig. 4 are now permitted all the way up to the top of the *p4gm* branch since inequality (9a) is fulfilled for all of the relevant non-disjoint geometric models of the input image data.

**Table VII:** Results of the hkaAICnorm MatLab script on the pattern with a large amount of added noise that underlies Fig. 6 for geometric model selection by G-AIC value ratios.

| Plane symmetry group to which the image data have been symmetrized | Sums of squared residuals of complex Fourier coefficients | Sums of squared residuals of Fourier coefficient amplitudes | Number of Fourier coefficients in the geometric model of the image data |
|---|---|---|---|
| *p2* | 0.0061 | None | 275 |
| *p1m1* | 1.5353 | 0.0039 | 271 |
| *p11m* | 1.5320 | 0.0039 | 271 |
| *p1g1* | 0.0069 | 0.0039 | 265 |
| *p11g* | 0.0078 | 0.0039 | 270 |
| *c1m1* | 0.0085 | 0.0041 | 269 |
| *c11m* | 0.0074 | 0.0041 | 269 |
| *p3* | 1.7565 | 1.2029 | 306 |
| *p2gg* | 0.0098 | 0.0039 | 264 |
| *c2mm* | 0.0115 | 0.0041 | 269 |
| *p4* | 0.0088 | 0.0028 | 276 |
| *p4mm* | 1.5876 | 0.0053 | 276 |
| *p4gm* | 0.0109 | 0.0051 | 266 |

**Table VIII:** Numerical values of ratios of sums of squared residuals of the complex Fourier coefficients of non-disjoint geometric models of the pattern with a large amount of added noise.

|  | Left hand side of (9a) | Right hand side of (9a) | Inequality fulfilled? |
|---|---|---|---|
| *p2gg* over *p2* | 1.6065574 | 2.04 | yes |
| *p2gg* over *p1g1* | 1.4202899 | 2.0037736 | yes |
| *p2gg* over *p11g* | 1.2564103 | 2.0222222 | yes |
| *c2mm* over *p2* | 1.8852459 | 2.0218182 | yes |
| *c2mm* over *c1m1* | 1.3529412 | 2.0 | yes |
| *c2mm* over *c11m* | 1.5540541 | 2.0 | yes |
| *p4* over *p2* | 1.442623 | 1.9963636 | yes |
| *p4mm* over *p4* | 180.4091 | 1.3333333 | no, blocking ascent |
| *p4gm* over *p4* | 1.2386364 | 1.3454106 | yes |
| *p4gm* over *p2gg* | 1.1122449 | 1.3308081 | yes |
| *p4gm* over *c2mm* | 0.947826 | 1.3370508 | yes |

It is interesting to check if this misidentification is consistent with the classification of the rather noisy pattern into the most likely projected Laue class as well. Table IX provides the basis for checking this out. Laue class *4* is, however, still identified by inequality (9a) as the one that minimizes the geometric AIC value. This might be due to 2D Laue class determinations being somewhat less susceptible to added noise.

Also there are many more calculations going into crystallographic symmetry classifications with respect to plane symmetry groups as compared to their counterparts for projected Laue classes. Rounding errors and approximations in the algorithms may therefore accumulate in the calculation for plane symmetry classifications more than for their counterparts for 2D Laue classes.

From the obvious crystallographic inconsistency that plane symmetry group *p4gm* and Laue class *4* have both been identified as K-L best representations of the rather noisy pattern in Fig. 6, one needs to conclude that the plane symmetry classification is probably too high and Fedorov type pseudosymmetries have been misinterpreted as genuine symmetries.

**Table IX:** Numerical values of the ratio of the sums of squared amplitude residuals of non-disjoint geometric models of the pattern with a large amount of added noise.

|  | Left hand side of inequality (9a) | Right hand side of inequality (9a) | Inequality fulfilled? |
|---|---|---|---|
| *4mm* (in *c2mm* setting) over *4* | 1.8928571 | 1.333333 | no, revealing an inconsistency |
| *4mm* (in *p2gg* setting) over *4* | 1.8214286 | 1.3454106 | no, revealing an inconsistency |
| *4mm* over *2mm* (in *p2gg* setting) | 1.3076923 | 1.3370508 | yes, as a result of pseudo-symmetry |
| *4mm* over *2mm* (in *c2mm* setting) | 1.2926829 | 1.3246592 | yes, as a result of pseudo-symmetry |

As mentioned above twice, crystallographic symmetry classification results as obtained in this section were to be expected and are in line with those of [10] for another series of synthetic patterns with and without added noise that feature pseudosymmetries. The conclusion from both studies must be that the information theory based classification methods work very well for small to moderate amounts of noise that is to a sufficient approximation Gaussian distributed. Methods that rely on the suppression of higher order terms in equation (3) must, however, fail if there is way too much Gaussian noise in a more or less 2D periodic pattern that is to be classified with respect to its crystallographic symmetries. Everything depends, of course, also on the relative complexity of a crystallographic pattern and the strength of its pseudosymmetries.

The classification failure is for the 2D crystal pattern in Fig. 6 not "catastrophic" as even when a misidentification is obtained for the most likely underlying plane symmetry group of the noisiest image, most human experts would have made the same mistake, at least at first sight. Because it is well known that Fedorov type pseudosymmetries are not rare [9] in nature, one needs to be extra careful with the crystallographic processing of very noisy images in order not to misinterpret noise as structural information.

In the following subsection, the noisy pattern of Fig. 5 is symmetrized to the plane symmetry group that our analyses indicated. We also symmetrize the very noisy pattern of Figure 6 to plane symmetry group *p4gm*, although our analysis indicated that there was a problem.

## C. Results of crystallographic image processing of the two noisy patterns of the analyzed image series

In order to demonstrate the benefits of the crystallographic image processing procedure, the classification results of the genuine plane symmetries of the noisy patterns in Fig. 5 and 6 are now used to boost the signal to noise ratio in these two crystal patterns.

Figure 7 shows approximately 2.2 unit cells of the *p4* symmetrized pattern of Fig. 5. The conspicuous bright bow ties in Fig. 7 feature now site symmetry *2* as exactly as this is possible for a real-world entity that has been derived from another real-world entity by the employed algorithmic procedure. (As a reminder: in Figs. 1 to 3 and 5, the bow ties feature point symmetry group *2* to a very good approximation. The entry for the *p2* symmetrized model of the noisy pattern in Fig. 5 features in Table IV the lowest sum of squared residuals of the complex Fourier coefficients, making this plane symmetry group the anchoring group.)

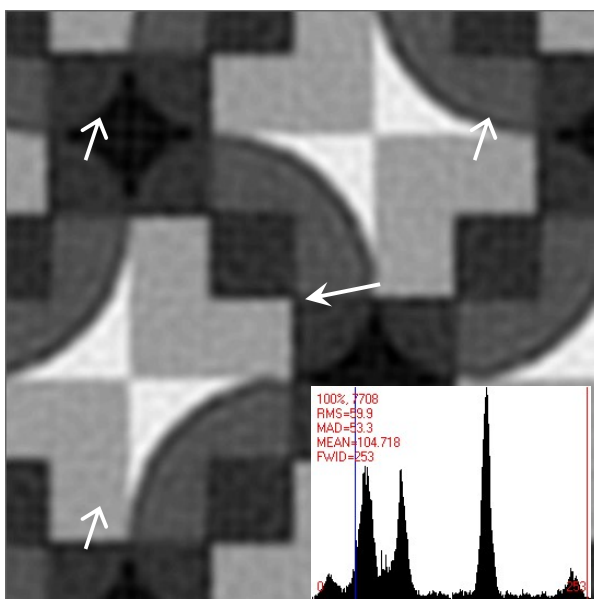

**Fig. 7.** Approximately 2.2 unit cells of the moderately noisy pattern of Fig. 5 after crystallographic image processing with histogram as inset. Minor artefacts that are due to the crystallographic image processing are marked by arrows of two different shapes. The reduction of the Abbe resolution, the symmetrization of Fourier ripples as well as the symmetrization of "remains" of the generalized noise to site symmetry group caused these artifacts.

Note by a visual comparison between the patterns in Figs. 5 and 7 how much of the added noise has been removed by the crystallographic image processing. This becomes also clear by a direct comparison of the histogram insets of both figures. A comparison of the histogram insets of Figs. 7 and 1 reveals the effect of the crystallographic image processing as well. Note that the overall contrast in Fig. 7 is lower than in Fig. 1.

Very small four-fold artifacts of the crystallographic image processing are marked in Fig. 7 by white arrows with two different shapes. One such translation periodic artifact is, for example, the imposing of exact site symmetry *4* on the centers of the dark rounded diamonds and the faint narrow-straight crosses[11] inside these diamonds. (This leads to an enhanced visibility of the narrow-straight crosses as compared to their counterparts in Fig. 1.)

The very low-contrast four-fold "feature" inside the curved dark diamond at fractional unit cell coordinate ½,½, is marked by an arrow with a different shape. It originated as well from the symmetrization of remains of the noise at the positions of the structure-bearing Fourier coefficients by the crystallographic image processing.

Analogously, note that the "bright bow ties" (at unit cell coordinates ½,0, ½,0, ½,1, and 1,½) are not homogenously bright as they were in Fig. 1. They feature instead some form of a "fine structure" that has been enforced to display the symmetry of two-fold rotation points. That fine structure is due to the symmetrization of a combination of Fourier ripples, as visible in Fig. 2, and remains of the noise.

All of these undesired features are, however, small prices to pay in the opinion of the author for a significant reduction in the noise level and the associated increase in the intrinsic image quality by means of the crystallographic processing of a noisy image.

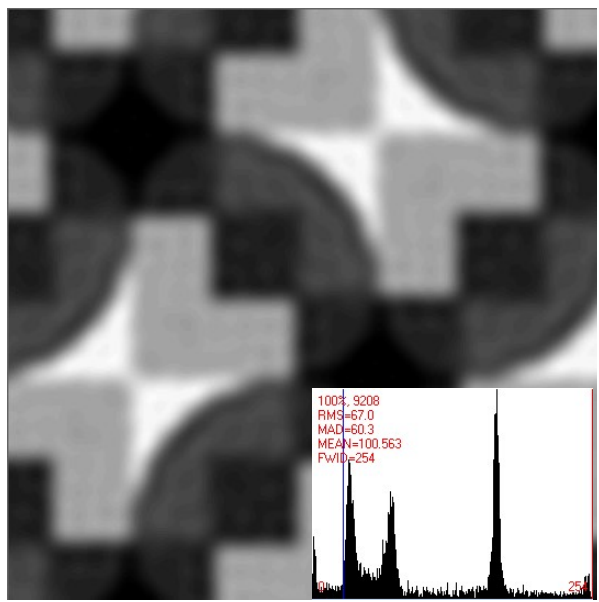

**Fig. 8.** Approximately 2.2 unit cells of the rather noisy pattern of Fig. 6 after crystallographic image processing with histogram as inset. The blue and red thin lines and descriptive annotations in red ink in the histogram are provided by the CRISP program. Note the reduction in contrast and Abbe resolution with respect to both the images in Figs. 1 and 6.

Essentially the same can be said about the crystallographically processed version of the very noisy pattern in Fig. 6. The contrast in the crystallographically processed version of this pattern in Fig. 8 is even lower than in Fig. 7.

---

[11] The structural origin of these narrow-straight crosses is the joining of four identical photos of the original square painted ceramic tile, Fig. A-7. These crosses are clearly visible in Fig. A-8 for arrangements of four tiles that feature a central four-fold rotation point.

The contrast reduction is mainly a consequence of using a smaller number of symmetrized complex Fourier coefficients for the transformation back into direct space (reduced Abbe resolution). The faint narrow-straight cross artifacts within the dark rounded diamonds from Fig. 7 have apparently disappeared in Fig. 8. Similarly the bright bow ties in that figure appear to have lost much of their fine structure.

Because plane symmetry group *p4gm* has been enforced on the very noisy image in Fig. 6, strong Fedorov pseudo-symmetries have been rendered visibly indistinguishable from genuine symmetries in direct space. The conspicuous bow ties feature in Fig. 8, therefore, point symmetry *2mm*, although the corresponding site symmetry in the undisturbed 2D crystal pattern was at best point group *2*, as most clearly visible in Figs. 2 and 3. Noise in the image has, thus, been misinterpreted as structure as part of the crystallographic image processing.

The increased narrowness of the peaks in the histogram inset of Fig. 8 with respect to their counterparts in the histogram inset of Fig. 7 is due to averaging over twice as many (wrongly identified) asymmetric units in the crystallographic image processing. This wrongful averaging (and Fourier back-transforming into direct space) created sites in the unit cells that feature now point symmetry group *2mm* at the fractional unit cell coordinates ½,0, 0,½, ½,1, and 1,½, compare to Fig. 2.

Nevertheless, the suppression of the noise in both of the noisy patterns is quite impressive when judged from the histogram insets in Figs. 5 and 6. Again, scanning probe microscopists should take notice of this fact as crystallographic image processing on the basis of objective crystallographic symmetry classifications is now available to them as well. They need, however, to be wary of Fedorov type pseudosymmetries that are easily misinterpreted as genuine symmetries when noise levels are high.

The large amount of added noise pattern, Fig. 6, was crystallographically processed in plane symmetry group *p4gm*, Fig. 8, although the Laue class classification identified a problem with the *p4gm* classification that is caused by the large amount of added noise. This was done here for the sake of a demonstration of what happens when one symmetrizes a more or less 2D periodic pattern to a plane symmetry group that is not crystallographically consistent with the corresponding projected 2D Laue class classification by the information theory based method.

Scanning probe microscopists should heed the advice that noisy images are only to be symmetrized to plane symmetry groups that are crystallographic consistent with the Laue class classification of a more or less 2D periodic image.

## V. COMPARISONS OF OUR RESULTS TO SUGGESTIONS BY THE CRISP PROGRAM

The objectively obtained crystallographic symmetry classification results of the previous section are summed up in Table X and are now compared to the results of a traditional classification with the electron crystallography program CRISP, Table XI.

It is clear from Table XI that the CRISP suggestions do not make distinctions between genuine symmetries and Fedorov type pseudosymmetries. As mentioned above, most human experts would most likely have classified all three images of the series as belonging to plane symmetry group *p4gm* (at least at first sight) because it would not occur to them that such distinctions might be necessary although it is well known that Fedorov type pseudosymmetries are not rare in nature [9]. As the analyses in the preceding sections demonstrate, *p4gm* classifications for the noise-free and large amount of added noise pattern constitute overestimations of the plane symmetry that is genuinely there, i.e. *p4*, due to the artist's graphic handiwork.

**Table X:** Plane symmetry and Laue class classifications of the analyzed set of patterns by the author's methods.

| Pattern | Plane symmetry group | Laue class |
|---|---|---|
| Free of added noise, Fig. 1 | *p4*, with strong *p1g1*, *p11g*, *c1m1*, *c11m*, and somewhat weaker *p2gg*, *c2mm*, *p4gm* Fedorov type pseudosymmetries | *4*, *2mm* and *4mm* pseudo-symmetries |
| Moderate amount of added noise, Fig. 5 | *p4*, with strong *p1g1*, *p11g*, *c1m1*, *c11m*, *p2gg*, and somewhat weaker *c2mm*, *p4gm* Fedorov type pseudosymmetries | *4*, *2mm* and *4mm* pseudo-symmetries |
| Large amount of added noise, Fig. 6 | *p4*, all Fedorov type pseudosymmetries at the plane symmetry group level were misidentified as genuine symmetries, but the identification of point symmetry *4* as the anchoring Laue class revealed their true nature and confirmed *p4* as the crystallographically consistent plane symmetry group classification | *4*, *2mm* and *4mm* pseudo-symmetries |

**Table XI:** CRISP program suggestions for the plane symmetry classifications of the analyzed series of patterns.

| Pattern | Plane symmetry group |
|---|---|
| Free of added noise, Fig. 1 | *p4gm* |
| Moderate amount of added noise, Fig. 5 | *p2gg* |
| Large amount of added noise, Fig. 6 | *p4gm* |

One has to conclude that two of CRISP's three suggestions are about as accurate as most human classifiers would be by visual inspection, at least at first sight. The program's internal thresholds for the interpretation of the CRISP residuals that enabled its plane symmetry estimations must, therefore, be well adjusted. The great popularity of this Windows-based computer program with the electron crystallography community is, therefore, well justified. It has been shown elsewhere that the CRISP program is more accurate in the extraction of the lattice parameters of more or less 2D periodic images than two of its competitors [8].

The *p2gg* classification by CRISP for the modest amount of added noise pattern, Fig. 5, is consistent with the bright bow tie features having a site symmetry that is no higher than point symmetry group *2*, as clearly revealed in Figs. 2 and 3. It is also consistent with the CRISP derived lattice parameter set of *a* = 97.1 pixels, $b$ = 97.0 pixels, and γ = 90.0° for the 2D crystal pattern in Fig. 5. The difference in the magnitude of the unit vectors is probably not significant based on what has been shown in [8].

As one can interactively test adjacent image areas for their CRISP program suggestions', one can not only assess the accuracy of the program's classification suggestions, but also their precision. It was found that adjacent areas in both the noise-free and moderate amount of noise added image resulted in either *p4gm* or *p2gg* classifications with CRISP. The *p4gm* suggestion by CRISP for the noisiest image did, however, not change with the selected image regions.

At least the noise-free image in the series should be homogenous so that all adjacent image areas should be classified as featuring the same plane symmetry. One has to note that a large amount of calculations goes into a plane symmetry classification so that the CRISP residuals for different geometric models of the input image data are indeed slightly different for each different image region.

Crystallographic symmetry classifications with the CRISP program rely in practice heavily on visual comparisons between the translation averaged (Fourier filtered) and differently symmetrized versions of the input image data by an expert practitioner of electron crystallography. Faced with a *p2gg* classification and a 2D Bravais lattice that is almost of the square type by CRISP (as obtained for the moderate amount of added noise image), most electron crystallographers would probably have simply overwritten that suggestion after visual inspections and concluded that the plane symmetry groups is *p4gm*, based on a square unit cell, and discounted the possibility of a metric specialization.

Using the author's information theory based methods, no *visual* comparisons between the translation averaged and differently symmetrized versions of the input image data are necessary. Crystallographic symmetry classifications can, therefore, be made without human supervision, but under the currently necessary assumption that there is indeed more than translation symmetry in a noisy image.

Therefore, the user no longer needs to be an electron crystallographer to employ crystallographic image processing techniques. This fact allows sufficiently well resolved more or less 2D periodic images from a wide range of crystalline samples that are recorded with different types of microscopes to be processed crystallographically. Previous successes in the crystallographic processing of images from scanning tunneling and atomic force microscopes are quoted and shown in [25,29].

## VI. SUMMARY AND CONCLUSIONS

Information theory based crystallographic symmetry classification methods for plane symmetry groups and projected Laue classes have been demonstrated on a series of three synthetic crystal patterns. The classifications were for the two noisy patterns complemented by the showing of the corresponding images and their histograms before and after their crystallographic processing. Note that these pairs of images needed to be shown in this paper for demonstration purposes, but crystallographic image processing by the new method can proceed without prior visual inspections of such images by human beings.

It is concluded that the information theory based classification methods are statistically sound and superior to all other existing methods, including the visual insights of human expert classifiers as far as their accuracy and precision at first sight is concerned. Information theory based methods should be developed for crystallographic symmetry classifications and quantifications in three spatial dimensions as there is also subjectivity[12] in the current practice of single crystal X-ray and neutron crystallography.

The detection of noncrystallographic symmetries (as defined in the introduction as being incompatible with translation symmetry) is beyond the scope of the demonstrated methods and there are no plans by this author to try to tackle that kind of problem.

**Acknowledgments**

The current members of Portland State University's Nano-Crystallography Group, Regan Garner, Choomno Moos, Grayson Kolar, Gabriel Eng, Noah Allen, and Lukas von Koch are thanked for critical proofreads of the manuscript. Regan Garner is also thanked for the graphs in Figs. 4, A-1, and A-2. Professor Eva Knoll of the Department of Mathematics of the University of Quebec at Montreal is thanked for a critical proofread of the manuscript, enlightening discussions per email, and the final appendix. Professor emeritus Emil Makovicky of the Department of Geosciences and Natural Resource Management of the University of Copenhagen is thanked for comments and stimulating discussions.

---

[12] In every single crystal X-ray or neutron diffraction based determination of an unknown crystal structure, one needs to assign a space group in which the subjectively most reasonable model for the structure is to be refined. Information theory, as defined in footnote 2, is partly about the selection of the model for experimental data that is statistically/objectively most justified by the data itself. Since the experimental data is in diffraction based crystallography of a geometric nature, a geometric form of information theory such as the one by Kenichi Kanatani is applicable.

When the symmetry classification (and quantification) methods of this paper have been generalized to three spatial dimensions, Walter C. Hamilton's well known significance tests of crystallographic R-values after refinements into non-disjoint space groups (W. C. Hamilton, Acta Cryst. 18, 502–510, 1965) could be considered superseded. This is because they have been set up as null-hypothesis tests. Information theory is widely considered to offer a superior alternative to null-hypothesis testing, see D. R. Anderson, Model Based Inference in the Life Science, A Primer on Evidence, Springer, 2008, for a gentle introduction on how to bring more objectivity to scientific studies.

# Appendices

## Appendix A: Hierarchy tree of the crystallographic 2D point groups that are projected Laue classes

Figure A-1 presents the hierarchy tree of the 2D crystallographic point groups that are projected Laue classes. The two ordering principles in this graph are the order of the point group/Laue class on the left and right hand side, increasing from the bottom to the top, and the maximal subgroup and minimal supergroup relationships. Maximal subgroups are connected to their minimal supergroups by arrows. The ratios of the sums of squared residuals of the amplitudes of the structure-bearing Fourier coefficients for climbing up from a lower level of the hierarchy to a higher level that are permitted by inequality (9b) are also given in this figure.

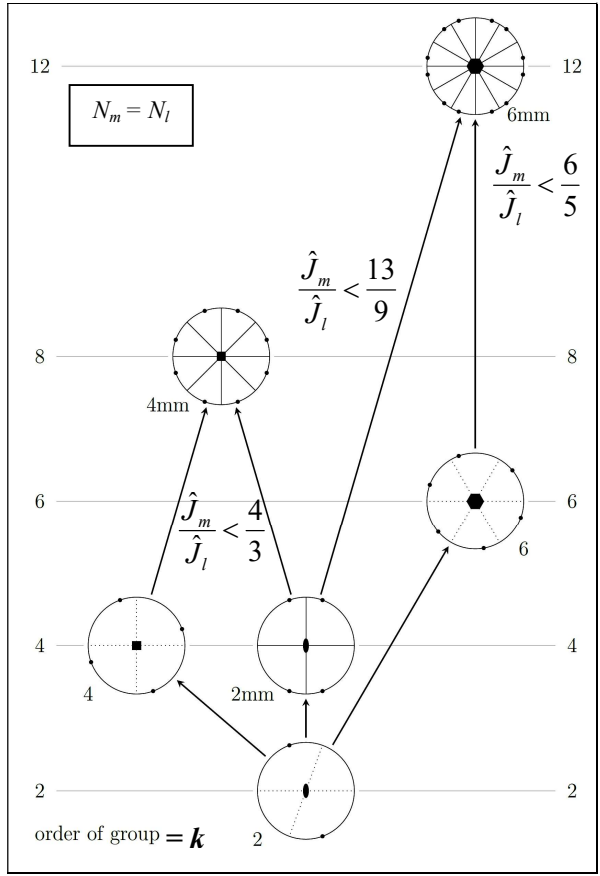

**Fig. A-1:** Hierarchy tree of the 2D point symmetry groups that are projected Laue classes with ratios of sums of squared Fourier coefficient amplitude residuals as insets. These ratios are valid for equal numbers of structure bearing Fourier coefficients of geometric models and apply to climbing up from a maximal subgroup at a $k_l$ level to its minimal supergroup at the $k_m$ level. Maximal subgroups are connected to their minimal supergroups by arrows.

Table A-1 lists the crystallographically consistent plane symmetry groups for each projected Laue class. The projected Laue class can be inferred for each plane symmetry group entry in [1,2] from the Patterson symmetry symbol by deleting the leading letter for the type of the unit cell.

Table A-1 projected Laue classes and their crystallographically consistent plane symmetry groups in short Herman-Mauguin notation.

| Projected (2D) Laue class | Plane symmetry group |
|---|---|
| 2 | p1, p2 |
| 2mm | pm, pg, cm, p2mm, p2mg, p2gg, c2mm |
| 4 | p4 |
| 4mm | p4mm, p4gm |
| 6 | p3, p6 |
| 6mm | p3m1, p31m, p6mm |

## Appendix B: Objective symmetry classifications from electron diffraction patterns from crystals

Objective symmetry classifications of electron diffraction patterns have a great future in connection with scanning transmission electron microscopy (STEM) using fast-readout pixelated direct-electron detectors. This relatively new technique is often referred to by the acronym 4D-STEM as 2D diffraction patterns[13] are recorded sequentially over a 2D grid of (direct-space) probe positions using a very fine electron probe [40].

A novel 4D-STEM imaging contrast mode that relies on the variation of the point symmetry of electron diffraction patterns with electron probe position changes on the order of a few tens of picometers was recently demonstrated [41]. The acronym *S*-STEM was introduced in that paper for this new scanning probe imaging technique, where the leading letter *S* stands for symmetry. A serious weakness in [41] is not to rely on a method for the classification of experimental electron diffraction patterns from crystals that deals objectively with symmetry inclusion relationships between the projected crystallographic point groups.

Such a classification method is, however, straightforwardly extrapolated from what is described above in this paper. For strictly kinematic electron diffraction conditions, one would have to use the hierarchy tree of the projected Laue classes, Fig. 1A, and the equations above. This is because Friedel's law applies per definition to such a scenario (as it often does in X-ray diffraction experiments). The electron diffraction patterns would then be strictly centrosymmetric about the position of the primary beam.

Most electron diffraction patterns for acceleration voltages of 100 to 300 keV feature, however, violations of Friedel's law [14,15,42]. For the classification of such electron diffraction patterns, the 2D Laue class hierarchy tree of Fig. A-1 needs be replaced by one which contains all 10 of the crystallographic 2D point groups. The same equations and inequalities as above apply for the ascent from lower symmetric point symmetry groups (that are maximal subgroups) to higher symmetric point symmetry groups (that are their minimal supergroups) on the basis of the ratios of the sums of reflection spot residuals in that other (not shown) hierarchy tree.

---

[13] As this electron probe is formed by a lens, one can refer to the recorded diffraction patterns as "images" of the probe-sample interaction. The recorded 4D-STEM dataset contains a very large amount of information about the local electron probe interaction with a material sample.

Violations of Friedel's law can be reduced by precessing the primary electron beam [42] close to precise zone axis orientations and transmitting the thinnest possible sample area [14,15]. Precession electron diffraction results also in more diffraction spots being present in spot electron diffraction pattern [42]. The symmetry of precession electron diffraction patterns [43] from several well chosen zone axes of highly symmetric crystals allows for 3D point symmetry reconstructions and facilitates space group determinations [44].

All of these features are useful for both electron crystallography [45] and structural fingerprinting [42,46]. Many diffraction based electron crystallography techniques [45] have so far avoided highly symmetric zone axis orientations of the crystals that are to be analyzed because dynamical diffraction effects are particularly prominent for such orientations.

When energy filtered electron diffraction patterns are available, one will be able to correct for dynamical electron diffraction effects [47] (and subsequently solve the lost structure factor phase problem of diffraction based crystallography as recently demonstrated on synthetic data). One will no longer have to rely on sufficiently thin crystals for electron crystallography on the basis of a multiple zone axis diffraction patterns [48].

The removal of dynamical electron diffraction effects will also benefit electron crystallography studies on the basis of STEM-through-focus- series images [49]. On the basis of such images [49] and by using a pixelated direct electron detector [40], one could proceed analogously to [23] and build up an objective plane symmetry group classification from objectively identified site/point symmetries [7,10,24,25]. As in any other kind of scattering-based crystallography, objectively determined 2D point and plane symmetries will facilitate the refinement of crystal structures that were solved from electron diffraction or imaging data.

It is also possible to expand the above described objective crystallographic symmetry classification methods to the point symmetries of quasicrystal. An amendment of the projected Laue class (centrosymmetric point group symmetry) hierarchy tree of Fig. A-1 would again be necessary.

A binary type classification method for distinguishing between quasicrystals and their rational approximants on the basis of the amplitude maps of the discrete Fourier transforms of sufficiently well resolved images from a crystal or quasicrystal is mentioned in [25]. Figure A-2 provides the projected Laue class hierarchy tree for such binary classifications. The ratios of the sums of the squared residuals of the amplitudes of the structure-bearing Fourier coefficients for objective classifications using the equations and inequalities above are given in this graph as well.

For electron diffraction spot patterns from quasicrystals in selected zone axis orientations, the hierarchy tree of Fig. A-2 would need to be amended by the crystallographic 2D point groups (as mentioned above for electron diffraction spot patterns from crystals).

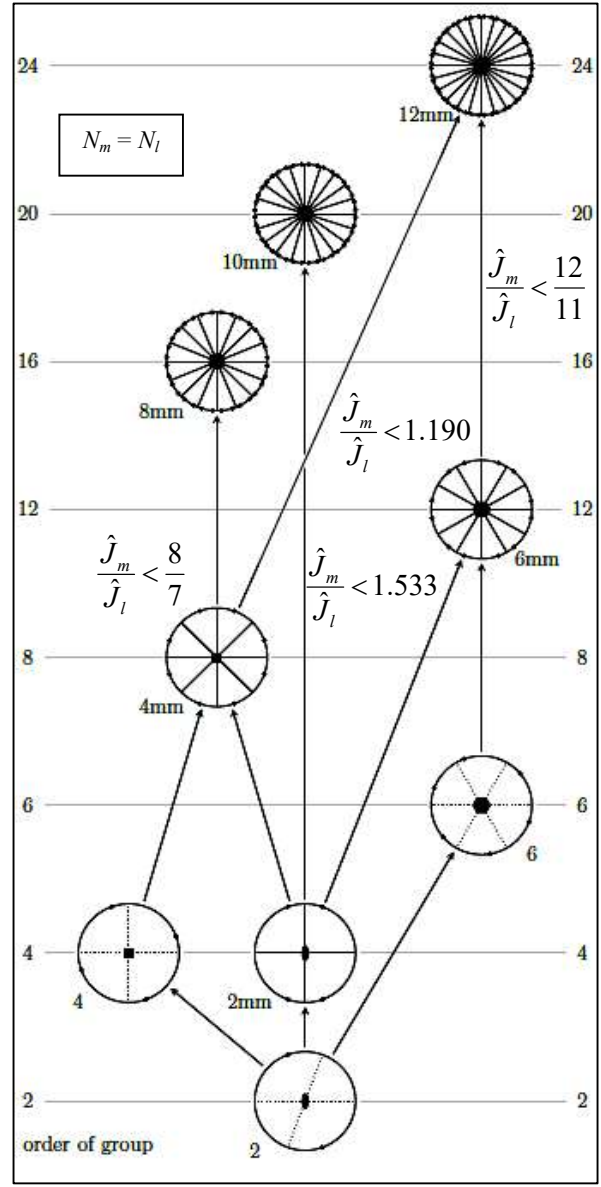

**Fig. A-2:** Hierarchy tree of the 2D point symmetry groups that are projected (2D) Laue classes of selected quasicrystals and their rational approximants. The ratios of the sums of the squared amplitude residuals are valid for equal numbers of structure bearing Fourier coefficients of geometric models and apply to ascents from a maximal subgroup at a $k_l$ level to its minimal supergroup at the $k_m$ level. Maximal subgroups are connected to their minimal supergroups by arrows.

*Appendix C: Ad-hoc confidence levels for classifications into minimal supergroups*

Based on Kanatani's information content ratio equation [27], ad hoc defined confidence levels for the model selections in favor of the more symmetric/restricted geometric model can, for the special case $N_m = N_l$, be straightforwardly defined. For two non-disjoint geometric models for which inequality 9b is fulfilled one obtains:

$$K = \sqrt{\frac{1 - \frac{1}{k_l}}{1 + \frac{1}{?}} \left( \frac{\hat{J}_m}{\hat{J}_l} + \frac{\frac{2}{k_m}}{1 - \frac{1}{?}} \right)} \quad \text{(A-1)},$$

The critical value for $K$ is obtained for inserting the condition

$$\frac{\widehat{J_m}}{\widehat{J_l}} = 1 \quad (A-2),$$

into (A-1) so that

$$K_{critical} = \sqrt{\frac{k_m - \frac{k_m}{k_l} +}{k_m + \frac{k_m}{l}}} \quad (A-3)$$

results.

Obviously, $K \geq K_{critical}$ is valid as the ratio of the two sums of squared residuals ranges from unity (A-2) to a constant value that is larger than unity and depends on the particular combination of $k_m$ and $k_l$ in inequality (9b).

When the ratio of the squared residuals is unity (as in equation (A-2), one has 100 % confidence in choosing the more symmetric model over the less symmetric model. Both models fit the input image data equally well in that special case, which will in practice only be obtained for noise-free mathematical idealizations of real world images, perfect geometric models, and with a perfectly accurate algorithm.

When inequality (9b) is not fulfilled, one has zero confidence in the selection of the more constrained model over its less symmetric counterpart. This is formalized by the confidence level in identifying a minimal supergroup

$$C_m = \frac{1 - K}{1 \quad K}(1 \quad (A-4),$$

which takes on values between zero and 100 % as a function of the ratio of the sums of squared residuals that allow for climbing ups from lower levels of the symmetry hierarchy to higher levels. It makes sense to define an average confidence level for a transition from all maximal subgroups to their joint minimal supergroup. For small symmetry breakings by each individual group and low-noise data, this average confidence level can be rather high.

The confidence level becomes negative and devoid of its usual meaning when inequality (9b) is not fulfilled. This is a result of the ratio of the squared residuals in A-1 becoming so large that K becomes larger than 1.

Pseudosymmetries are identified by high confidence levels for transitions that are not based on the anchoring group.

*Appendix D: Crystallographic image processing as a form of computational imaging*

Computational imaging is in [20] defined as the union of digital data recording and processing that results in a much improved image of some physical interaction between some sample with probing particles or forces in any dimension. The concept of an "indirect" imaging system is also defined in [20], whereby the image-data processing is an indispensable part of the imaging process (as it is, for example, in tomography, holography, ptychography, super-resolution fluorescence microscopy, and image-based electron crystallography).

The spatial resolution of the physical apparatus used for the initial data collection part of indirect imaging, e.g. the pixel size of a digital detector and a microscope's magnification, determines no longer the spatial resolution of a study in computational and indirect imaging. Of central importance in computational imaging is the concept of the dimensionless intrinsic image quality [20]

$$Q_{as\_recorded} = \frac{SNR}{SR_{PSF}^{d/2} \cdot \sqrt{F_{in}}} \quad (A-5),$$

where $SNR$ stands for the signal to noise ratio, $SR$ is the spatial resolution as defined by the widths of the instrumental point spread function ($PSF$), and $d$ is the dimension of the image (equal to two for the images in this study). The quantity $F_{in}$ in this equation is the imaging particle fluence, expressed as the number of registered quanta (*in*put) per image-unit area (or generally per $d$-dimensional volume) of a digital image. In computational imaging, $SR_{PSF}$ and $SNR$ can be traded-off each other and the intrinsic image quality of the as recorded image is on the order of unity [20].

Note the duality of the signal-to-noise ratio and spatial resolution in equation A-5 for a fixed imaging particle fluence. While the image size, its bandwidth, and signal-to-noise ratio (or, equivalently, the structural (spatial/angular) resolution and the $SNR$) can all be traded-off against each other, their product cannot exceed the number of image forming quanta. The maximum *"information capacity"* of a computational imaging system is limited by the number of quanta used in the formation of the image [50]. Additional information is often available in the form of prior knowledge about certain aspects of the imaged sample.

The intrinsic image quality can be significantly increased by the application of such prior knowledge. In crystallographic image processing, this knowledge comes in two forms: (*i*) the sample is known to be translation periodic in a 2D projection and (*ii*), the projection of the space group symmetry features a plane symmetry higher than *p1*. When a quasicrystal is imaged with a computational imaging system, it is known that the amplitude map of the discrete Fourier transform of its image consists of discrete Bragg spots that feature a (2D) projected Laue class that is non-crystallographic and typical of a quasicrystal, see Fig. A-2.)

The total area of the image on which the imaging particles impinged is taken in this paper to be composed of $K$ crystallographic unit cells. Since Fourier filtering is analogous to translation averaging over all noisy unit cells of the image with subsequent removal of all non-structure bearing Fourier coefficients, a factor of the square root of $K$ enters the nominator of equation (A-5) when the averaged particle fluence per unit cell area (instead of the total image area) is used.

$$Q_{Fourier\_filtered} = \frac{SNR_{periodic\_structure} \cdot \sqrt{K}}{SR_{PSF} \cdot \sqrt{average\_F_{in}}} \quad (A-6).$$

The effect of the translation averaging by Fourier filtering can, therefore, be considered as a significant boost to the signal to noise ratio [30] that affects the intrinsic image quality positively. Information on the real crystal structure is, however, removed by Fourier filtering as recovering that kind of information was considered outside of the goal of the investigation. This information is filtered out as if it were just noise.

A reduction of the widths of the PSF is equivalent to an increase of both the structural (spatial/angular) resolution and the intrinsic image quality. Reductions of the widths of the PSF can be obtained by deconvolutions [20].

The crystallographic processing of a translation averaged image is in essence a deconvolution/removal of the "position- dependent point-symmetry-smearing" part [20] of the PSF from said image in Fourier space [51]. The rest of the computational imaging part of the "complete PSF" [20] remains as a feature of the digital image after this "site/point-symmetry-restoring partial deconvolution."

This kind of deconvolution can be considered as being equivalent to a posterior symmetrizing/shaping of the scanning probe tip in scanning probe microscopy. The partial image deconvolution by means of crystallographic image processing (CIP) is formally equivalent to the averaging over the $k$ noisy asymmetric units of the translation-averaged unit cell. No simple formula captures the increase of the intrinsic image quality with the crystallographic processing of an image

$$Q_{CIP} = CIP_{factor}(k, motif...) \cdot Q_{Fourier\_filtered} \quad \text{(A-7)},$$

because the "boost factor" depends on the specifics of the translation periodic motif, its underlying plane symmetry, the physical pixel size, the number of pixels per translation- averaged unit cell, and other measures of the physical-imaging instrument-part of the complete PSF. As Fig. 4 shows, the correctly averaged asymmetric unit can be up to $k_{max} = 12$ times smaller than the translation-averaged unit cell.

Since it is connected to the multiplicity of the general position ($k$), the boost factor in equation (A-7) should be proportional to the square root of that multiplicity. Digital images from higher symmetric crystals gain, thus, more from CIP that those from their lower symmetric counterparts.

Scanning probe microscopy of more or less 2D periodic images fell previously not under the strict definition of computational imaging [20] since it often ended with Fourier filtering and required the value judgment of a human being on what the most likely plane symmetry group may be. This has now changed as objective crystallographic symmetry classifications (that do not rely on such value judgments by human beings) were enabled by this author's recent work [7,10,24,25,29].

### Appendix E: Hypothetical climbing up scenarios in the hierarchy trees of translationengleiche plane symmetry groups and projected Laue classes

Let us assume for the sake of the argument that the geometric model with the lowest sum of squared residuals was the one that was symmetrized to plane symmetry group *p2* and Laue class *2*. This plane symmetry group features $k_l = 2$ in the plane symmetry hierarchy tree in Fig. 4.

We see that this plane symmetry group is in minimal supergroup relationships with plane symmetry groups *p2mg*, *p2gm*, *p2gg*, *p2mm*, *p4*, and *c2mm* at the $k_m = 4$ level, as well as with *p6* at the $k_m = 6$ level, Fig. 4. Let us further assume that the sums of squared residuals for plane symmetry groups *p1m1*, *p11m*, *c1m1*, and *c11m* are two to three orders of magnitude larger than for *p2*. This would mean that there are no mirror lines in the input image data and plane symmetry groups that contain mirror lines at the $k_m = 4$ level are to be ruled out as correct classification of the input image data.

Let us further assume that the sums of squared Fourier coefficient residuals for the models that were symmetrized to plane symmetry groups *p4* and *p6* are also two to three orders of magnitude larger than that for *p2*. This means that the input image data does not contain four-fold or six-fold rotation points. The sums of squared residuals for the models that have been symmetrized to *p1g1* and *p11g*, are, on the other hand, in our example rather low so that it seems possible that the corresponding symmetrized models of the input image data could combine with the *p2* symmetrized model to the model with plane symmetry *p2gg*. As long as inequality (9a) is fulfilled for all three cases *p2*, *p1g1*, and *p11g*, we can "climb up" from all three of the lower symmetric models at the $k_l = 2$ level to the model that has been symmetrized to plane symmetry group *p2gg* at the $k_m = 4$ level of Fig. 4.

Plane symmetry group *p2gg* is a maximal subgroup of *p4gm*, so that another test with inequality (9a) needs to be made if a further climbing up in the plane symmetry group hierarchy tree may be permitted. Let us assume this test is negative (as there are no four-fold rotation points in the input image data). The model that features plane symmetry *p2gg* is then the Kullback-Leibler best model, which allows for the best separation of the structural information in the input image data from the non-information/noise. It was arrived at without an estimate of the prevailing noise level.

We now make up another sake of argument scenario. Let us assume that the correct plane symmetry classification is *p2gg*, but that the geometric model of the input image data that was symmetrized to *p4* features also a rather low sum of squared residuals of the complex Fourier coefficients. Plane symmetry group *p4* is then identified as Fedorov type pseudosymmetry as it is disjoint from plane symmetry group *p2gg*, which is the highest genuine symmetry.

In a third sake of argument scenario, let us assume that the sum of squared residual of the complex Fourier coefficient of the model that has been symmetrized to plane symmetry group *p4* is only modestly larger than those for the *p1g1* and *p11g* models and, therefore, sufficiently low that one can not only climb up from *p2*, *p1g1*, and *p11g* to *p2gg* (and no further beyond), but also from *p2* to *p4* (but no further beyond). The image features then a strong Fedorov type pseudosymmetry.

When a *p4* pseudosymmetry of the Fedorov type in some input image data is very strong, the unit cell in the image will be nearly of the square type. If the $a = b$ lattice parameter condition for a square Bravais lattice type is fulfilled within error bars of their measurements, one speaks of such a *p4* pseudosymmetry as being simultaneously of the Fedorov type and a metric specialization. Key to this classification is the fact that plane symmetry *p4* is measurably more severely broken than *p2*, *p1g1*, and *p11g* in our third sake of the argument scenario.

The *p4* symmetrized model of the input image data features 2D Laue class *4*. For consistency of the plane symmetry classification for our last scenario, the sum of squared residuals of the Fourier coefficient amplitudes for Laue class *4* needs to be higher than that for Laue class *2mm*. See the first appendix for the hierarchy tree of the projected Laue classes. The variables $k_m$ and $k_l$ refer in the context of classifications into 2D Laue classes to the order of the crystallographic 2D point groups on which these classes are based. All of the equations and inequalities above are valid with the modified meaning of $k_m$ and $k_l$ when information theory based classifications into Laue classes are made.

Projected Laue class *2mm* is the one that is consistent with all three geometric models of the input image data that contain glide lines in our third scenario, which have been identified as genuine symmetries of the input image data. At the center of the amplitude map of the discrete Fourier transform of the input image data, there can, therefore, only be a four-fold rotation pseudosymmetry point superimposed on a genuine two-fold rotation point that is located at the intersection of two mirror lines.

*Appendix F: Comments on alternative computational symmetry and machine learning approaches to crystallographic symmetry classifications*

Individual members of the computational symmetry community [21] are to be credited with realizing the importance of Kanatani's geometric form of information theory [26-28] for their field roughly five years earlier than this author did. As mentioned above, this statistical theory deals objectively with symmetry inclusion relations and ensures that the statistically most likely symmetry can be reliably identified in noisy 2D periodic images.

The works of the computational symmetry community are concerned with direct space analyses and sometimes lack crystallographic rigor (as evident by a display of ignorance about the concepts of standard plane symmetry group origin choices and the importance of site symmetries [1,2]). Maybe this is one of the reasons that only two studies exist, to the best of this author's knowledge, where a G-AIC was used in the classification of a time-periodic series of 2D images in direct space. These two studies ignored the standard origin choices [52] for frieze groups and their results are somewhat questionable, as discussed in appendix E of [7]. If the two above-mentioned periodic time series had been analyzed in Fourier space, the alignment of the raw data with the symmetrized models for that data would have been trivial and better results could presumably have been obtained.

The conceptual difficulty in distinguishing between pseudosymmetries at the point/site level and genuine symmetries was noted in [22] by other computer scientists, but no solution to this problem was given. Since this difficulty is conceptual, it needs to be overcome by precisely defining what constitutes a genuine symmetry group, on the one hand, and a Fedorov type pseudosymmetry group, on the other hand. This has been done in this paper in line with the theory of crystallographic groups, whereby the least broken symmetry group acts as 'anchoring group' and anyone of its minimal supergroups is a genuine symmetry only if all of the other maximal subgroups are also passing the "climbing-up" test. Failed climbing up tests identify a pseudosymmetry in all those cases where the corresponding crystallographic groups feature also relatively low squared residuals of the complex Fourier coefficients.

Just as any other crystallographic symmetry classification study by the computational symmetry community, the method in [23] relies on arbitrarily set thresholds. Their method, is, however, quite unique in so far as it identifies the higher symmetric site/point symmetries and combines them to plane symmetry group classifications.

The preoccupation of the computational symmetry community with making classifications in direct space is from the perspective of natural scientists and engineers misguided. In science and engineering, one would always use Fourier methods when one is confronted with a problem that is periodic in space or time. When the sequence of periodic repeats is small, one would simply augment it with multiple copies of the original data in order to enable effective space or time "translation-averaging" by a discrete Fourier transform. One would, of course, also make sure that no artifacts were introduced into the data due to the procedure that was used for the augmentation[14].

Note that except for the two above mentioned studies of periodic time series, all of the so far reviewed [34] crystallographic symmetry classifications into plane symmetry groups, frieze groups, site/point symmetry groups, and 2D Bravais lattice types by the computational symmetry community involved internally programmed thresholds for automated interpretations, i.e. subjectivity disguised in computer code in other words. These classifications are, therefore, all subjective at a fundamental level in spite of having been under development for more than half of a century.

The computer scientist Xanxi Liu, who coined the term "computational symmetry" for the analysis of more or less 2D periodic patterns some 20 years ago, has recently started to use machine learning techniques for symmetry classifications of everyday objects from digital 2D images [53a]. This seems like a proper ending of the more than 50 years quest to find the best possible computer program for crystallographic symmetry classifications that combats the symmetry inclusion problem with subjectively set thresholds, which seems in hindsight somewhat ill-advised.

A few years later, Liu and her co-worker presented quite impressive results on the classification of synthetic 2D periodic patterns in direct space that were created out of random patches of noise [53b]. This kind of training data is probably optimal as its only common features are the systematically enforced crystallographic symmetry transformations of the Euclidian plane.

Their four-layer convolutional neural network performed within error bars about as well as Xanxi Liu and co-worker's analytical direct space approach [53c] with presumably very well adjusted thresholds from the year 2004. This result suggests that the efforts dedicated to trying to solve an image symmetry classification problem with a newly developed machine learning approach are not well justified when there is already a well performing analytical solution to the problem.

The minimal number of labeled training patterns per plane symmetry group for a high classification accuracy was determined to be 600. For such (600 times 17) an amount of training images, the classification accuracy ranged from 90.85% for plane symmetry group *p3* to 99.96% for plane symmetry *pg*. For a ten times larger training dataset, the respective classification accuracies for these two plane symmetry groups increased to 93.52% and 99.97%, respectively. A further eight-fold increase of the amount of training data changed these percentages only very marginally, whereby the classification accuracy for plane symmetry group *p3* actually dropped by 0.22 %.

---

[14] Weak superstructure reflections may appear in the discrete Fourier transform amplitude map if one had stitched several slightly different direct space periodic repeats together. One should then only use the main reflections, i.e. the Fourier coefficients that represent the average structure for crystallographic symmetry classifications. No superstructure reflections were observed in the amplitude map of the crystal pattern in Fig. 1 as all repeats in the original composite graphic work of art were identical, i.e. one physical tile represents one asymmetric unit and one quarter of the translations periodic unit cell, see Fig. A-7 and A-8.

Misclassifications were typically obtained more often for a supergroup of the correct group, e.g. *p6* instead of *p3*, than for a subgroup of that group, i.e. *p3* instead of *p6*. This means that Federov type pseudosymmetries were more often misclassified as genuine symmetries than genuinely present symmetries missed. The authors of [53b] are well aware of Kanatani's 1997 comments [26] on the symmetry inclusion problem and the unavoidable shortcomings of using any traditional measure of a minimized "symmetry distance" as basis of a classifier, as they quote him in the "Discussion and future work" section of their paper.

It will be interesting to see classification results for noisy images that are announced as one of the directions of their future work [53b]. As it was to be expected, their adding of symmetry breaking patches into unit cells in direct space resulted in misclassifications. This does not bode well for future applications of the machine learning system of [53b] to the crystallographic symmetry classification of noisy direct-space images. In this author's classification method [7,10,24,25,29], which works in Fourier space, such patches would effectively be removed by translation averaging.

Plane symmetry group classifications were also undertaken on the basis of the discrete Fourier transform coefficients of training images in [53b]. The accuracy of those classifications were, however, barely above random chance. This is probably the result of not having applied proper phase shifts on these coefficients so that they all refer to the crystallographically defined standard origin of the average unit cell (as it is done in this author's method.)

There is also a new breed of materials-data scientists who used machine learning systems for crystallographic symmetry classifications of direct and reciprocal space 2D images from crystals [54-56,58-64]. All of their classifications were performed in reciprocal/Fourier space, as this is the most sensible thing to do. Their general approach might at first sight appear to be semi-objective as no subjectively set *"thresholds for interpretation"* are programmed into the computer code. Machine learning classifications are (for contemporary systems) always functions of the sets of labeled images that were used in the training of such machines. Decisions on the composition and size of the training image set are obviously subjective.

When one relies for the training data sets on symmetry information about crystals that has been collected in the large crystallographic databases, one encounters the additional problem that the crystal structures of chemical compounds are extremely unevenly distributed over the various space groups. This is true for both inorganic [65] and organic materials [66,67].

While many inorganic crystals feature high symmetries, the opposite is true for organic crystals. Approximately one third of all organic compounds crystallize for example in space group $P2_1/c$. Approximately three quarters of all organic compounds crystallize in one of only five space groups. For inorganic compounds, there are only 24 space groups that feature more than 1 % of the (more than 100,000) entries in the 2006 edition [63] of the Inorganic Crystal Structure Database.

Individual more or less kinematic 2D diffraction patterns from real crystals and calculated amplitude maps of discrete Fourier transforms from more or less 2D periodic images probably do not contain enough symmetry information to arrive at unambiguous space group classifications by machine learning systems (in either 2D or 3D) even when complications due to symmetry inclusion relations are ignored. Note that there are only 11 Laue classes (centrosymmetric crystal classes and point groups) but 230 space groups in three dimensions, projecting to six 2D Laue classes and 17 plane symmetry groups (with 21 settings) [1,2]. Please note also that the comprehensive/systematic listing of the maximal subgroups of the space groups takes up over 300 pages in [68].

There are, thus, a lot of symmetry inclusion relations to be considered in 3D. In one dimension, there are only seven space groups. Their symmetry inclusion relationships have recently been exploited in order to increase the performance of a neural network [69] in a task other than a crystallographic symmetry classification.

The authors of [55] state in the abstract of their 2018 paper that the *"current"* crystallographic symmetry classification methods that are based on diffraction spot patterns "*require a user-specific threshold, and are unable to detect average symmetries in defective structures*". That statement was essentially correct at that time and is in line with the conclusions of this author [34] with respect to the vast majority of computer programs for crystallographic symmetry classifications that have ever been developed by the computational symmetry community [21-23,53c], materials scientists and crystallographers.

Classifications of kinematic 2D "diffraction fingerprints" into eight highly symmetric structural prototypes of mono-element crystals were made with a high accuracy in [55]. Each of the diffraction fingerprints consisted of the superposition of six simulated kinematic diffraction patterns for γ-rays (with a wavelength that is typical for transmission electron microscopy). It was convincingly demonstrated that the method of [55] is robust with respect to the presence of point defects in the structural prototype crystals that are to be classified.

This robustness should be beneficial when diffraction fingerprints for crystallographic symmetry classifications are eventually constructed out of experimental diffraction data. As it is very well known that electron diffraction is almost always highly dynamic in nature, there will be significant challenges to overcome when experimental data are to be classified. For a first demonstration of the effectiveness of their new method, synthetic data that were calculated in the kinematic approximation sufficed completely.

The authors of [55] are in their own words well aware of the fact that a "*two-dimensional diffraction fingerprint … is not unique across space groups*" and *"cannot represent non-centrosymmetric structures"*. Only classifications into a few centrosymmetric structural prototypes were, therefore, demonstrated. These structural prototypes were well chosen as they represent more than 80% of the crystal structures of the chemical elements.

Highly commendable is that the authors of [55] provide an on-line tutorial for their method in accordance with the principles of reproducible research. It is also highly commendable to put the referee reports on the originally submitted manuscript of their paper into the supporting material of [55]. Overall, this is as one would expect from a paper that has been published in the journal "nature communications" and has by now (middle of March 2023) garnered some 250 quotes on google scholar.

Returning to the direct quotes from the abstract of [55] a couple of paragraphs above, their method is indeed free of user-specified thresholds, robust, and delivers probabilistic classification results. In the absence of an infinitely large training dataset, there is, however, necessarily some subjectivity in the selection of any finite training dataset. Algorithmic objectify is, on the other hand, guaranteed by the usage of geometric Akaike Information Criteria in the author's information-theoretic methods [7,10,24,25,29] for crystallographic symmetry classifications in 2D.

The authors of [56] state that they have been inspired by [55]. Indeed, that paper is also about classifying synthetic diffraction spot patterns into a few highly symmetric structural prototypes and was published in an interdisciplinary chemistry journal with a high impact factor. Again, the training data relied on the kinematic diffraction approximation and used the same computer program as [55] for their generation.

The authors of [56] decided, however, to recast their diffraction pattern recognition task into a computer vision task. Their recasting was from a crystallographic viewpoint unscientific and, therefore, ill-advised for a paper that is addressed to the wider chemistry community.

The otherwise laudable intention was to capture visual "similarities" in the appearance of individual diffraction spot patterns. One of these similarities was the projections of 3D Bravais lattice types along both highly symmetric and lower symmetry zone axes. The defining projected symmetry of the rectangular-centered 2D Bravais lattice type, i.e. two basic lattice vectors of equal magnitude that form an angle of neither 120, 90, or 60 degrees, was, however, discarded by labeling the corresponding training patterns as belonging to the oblique type, i.e. the one that is devoid of all symmetry restrictions.

Another similarity that was considered relevant for the labeling of the synthetic diffraction spot patterns were noticeable intensity differences of groups of Bragg reflections from different structural prototypes that belonged to the same space group. This included the zero intensity of systematically extinct reflections. Confusing structural prototypes with space groups, the paper claims that *"the space group 227 may contain 3 effective space groups based on the SADPs considering the brightness of the diffraction spots"*. (SADPs stands there for selected area diffraction patterns.)

The third labeling relevant similarity was a concept that was referred to as "chirality". (From a scientific view-point, kinematic 2D diffraction patterns are never chiral in the scientific meaning of the word as they can only feature the point symmetry of a projected Laue class.) In aggregate the labels on the training data were partially non-crystallographic so that [56] does not represent progress beyond what has already been demonstrated in [55].

The authors of [56] suggest in earnest that traditional electron crystallography is on its way to become obsolete as a result of their work. They write for example that *"regrouping the"* simulated (kinematical) selected-area electron diffraction patters *"in accordance with similarities in 2D patterns would be useful to train both the machine and nonexpert human by eliminating the necessity to 'learn' crystallography in detail."* While their novel way of regrouping/labeling was essential to [56], there is obviously no need for human beings to take up a computer vision approach. This is precisely because the latter are not mere "correlation detection" machines and are predominantly interested in the cause and effect relationships that an analytic approach is able to elucidate.

On a positive note, the authors of [56] demonstrated awareness of both the symmetry inclusion problem and that many symmetries are lost in projections from 3D to 2D.

Whereas some kind of a replication of the results of previous studies, e.g. [55], is always appreciated, the shortcomings of [56] are too numerous to be ignored. These shortcomings make this study quite irrelevant as a predecessor of the real-world task of classifying diffraction spot data into structural prototypes with a neural network. Whereas there is not much to nothing of value for a chemist or crystallographer in [56], it was demonstrated that more hidden layers in the neural network do not always increase the classification accuracy.

The authors of [56] state that in order to incorporate experimental diffraction patterns into the training sets of their neutral network at some later point in time a *"possible approach would be the use of the image-to-image translation algorithm which can convert the experimental data into what appears to be the simulated data."* That approach is illustrated in [57] (as quoted in [56]) with "color translations" between images with zebras and horses in the foreground, see Fig. A-3.

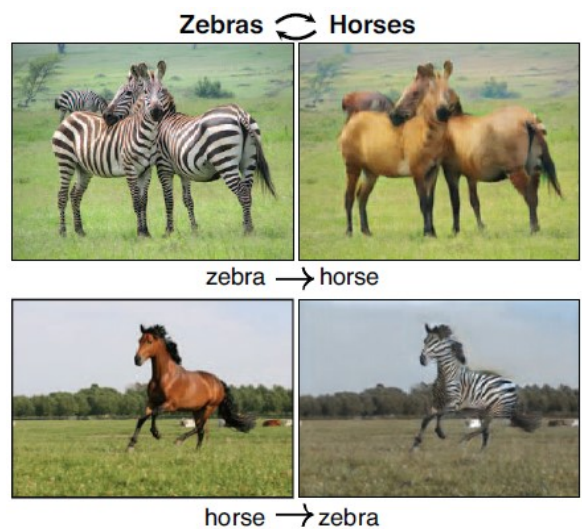

**Fig. A-3:** Illustration of the result of an unpaired image-to-image translation algorithm using cycle-consistent adversarial networks, from [57] (with permission). Note that the depiction of the "horras/zebses" in the upper right subfigure reveals, on closer inspection, stripes that are one of the defining characteristics of the zebras to their left (from which they were calcu/trans-lated).

The authors of [56] proposed (presumably in earnest) the usage of such a technique as source of simulated electron diffraction spot patterns that are derived from their experimental counterparts (which feature most likely dynamical diffraction effects). Unfortunately, there are no referee reports on the original manuscript of [56] in its on-line supporting material (as was the case for [55]).

One may wonder how any referee and the managing editor may have missed the wrong word choice in the second part of one of the introductory sentences *"…, generating geographic relations between the beam direction, sample orientation, and particular crystallographic planes."* (The authors of [56], all presumably being non-native speakers of English, are graciously excused for their choice of the word "geographic" instead of "geometric".)

Structural and symmetry information that has been lost by the recording of the outcome of a crystallographic diffraction experiment (or the extraction of Fourier transform amplitudes of a more or less 2D periodic image), i.e. the equivalent of collapsed wave functions, can probably not be recovered by contemporary "correlation detection machines". The goals of the study of [58] are, therefore, way too ambitious. (Using idiosyncratic neologisms for well established crystallographic concepts such as some of the geometric/morphological crystal classes [1,2] is also a shortcoming of [58].)

The authors of [58] claimed that they boosted the performance of their machine learning system by using predicted crystal systems as features for subsequent space group predictions. The usefulness of this strategy could, however, not been confirmed in a study on machine learning based crystal system and space group classifications on the basis of powder X-ray diffraction patterns [70]. This discrepancy could be due to the study in [58] being about symmetry in two dimensions (as projected from three dimensional crystals) and [70] being about 3D crystal symmetry projected into one dimension only. However, there is also a crystallographic-statistic reason that the "stacking strategy" of [58] is probably not optimal. Translational pseudosymmetries are definitively not rare in nature [71-74] and not all of the entries in the large crystallographic databases are correct [7,67].

It is not uncommon for non-crystallographers to mistakenly think that 2D lattice parameters which are within experimental error bars equal to each other, 90 or 120 degrees (60° as complement to 180°) allow for an assignment of noisy 2D periodic image data to higher symmetric 2D Bravais lattice types. The point group symmetry of a crystal structure restricts the metric properties of the translation lattice of any crystal structure (over restrictions on the components of the metric tensor) by the Neumann-Voigt-Minnigerode-Curie-principle [75,76]. In the "opposite direction", there is no principle that requires a triclinic crystal with a very strong translational pseudosymmetry or metric specialization [8] (that might be too strong to be identified as such in the unavoidable presence of experimental noise) to feature a metric symmetry higher than that of the triclinic Bravais lattice, i.e. $P\bar{1}$.

Correspondingly, a study of the distribution of organic crystal over the space groups (from the year 1983) found that *"depending on values used for 'reasonable' errors on the cell parameters, it was found that for 3 to 12 % of the compounds the metric symmetry exceeds the crystal symmetry reported by the authors in the original reference"* [67].

The study in [59] provides indirect support for the above presented hypothesis that there is probably insufficient symmetry information for the classification tasks at hand in individual amplitude maps of discrete Fourier transforms of digital images and more or less kinematic electron and X-ray diffraction spot patterns from crystals. The classification results of their neural network *improved* [59] when algorithmic "shape information" was *added* to the individual images of the training set. Adding information to the training set suggests to this author simply that there was not sufficient information there in the first place to obtain good classification results. The authors of [59] are, however, well aware of the difficulties that symmetry inclusion relation pose as they write in their abstract: *"deep learning (DL) … the DL-based identification of crystal symmetry suffers from a drastic drop in accuracy for problems involving classification into tens or hundreds of symmetry classes (e.g., up to 230 space groups), severely limiting its practical usage."*

Backscattered electron diffraction patterns feature Kikuchi lines and bands. They often cover a large field of view in reciprocal space so that some 3D structural information about a crystalline sample is contained in them. Crystallographic symmetry classifications on the basis of such patterns [60-62] by machine learning systems could, therefore, be superior in their accuracy to those from transmission electron diffraction spot patterns. A good strategy for the building up of a relevant training data set of experimental images is the elaboration of the experimental imaging parameters that affect the performance of their machine learning system [60,61] strongly [62].

More structural information than in essentially kinematic 2D electron diffraction patterns is encoded in dynamical convergent beam electron diffraction (CBED) patterns. This includes a limited, projection specific, amount of structural information in 3D and even some structure factor phase information when convergent beam electron diffraction disks overlap (and the crystals are sufficiently thick for dynamical electron diffraction effects to dominate). One may, however, remain skeptical if that additional amount of structural information may suffice for reliable space group determinations so that *"a universal classifier for crystallographic space groups"* [63] might eventually be created or remain a pipe dream.

Analytic methods of quantifying crystallographic symmetries in CBED patterns have, on the other hand, been developed and refined since the year 1985 [77-81]. The proposed symmetry deviation quantifiers of these studies are, however, not in a form that would allow them to become part of a geometric AIC.

Aspects of these analytic methods could, nevertheless, be adapted to work in conjunction with a novel crystal symmetry-based contrast mechanism for atomic resolution imaging [41]. That contrast mechanism uses so far exclusively a cross-correlation quantification technique of the shape of planar figures (with emphasis on the figure edges but without inherent crystallographic restrictions) [82]. As that symmetry classification technique originated in the computational symmetry community, it treats Bragg diffracted electron intensities and diffusely scattered background intensities in the same manner. Because the Bragg intensities carry the directly quantifiable bulk of the interpretable symmetry information from the crystal, there is room for improvement of the contrast mechanism of [41]. This crystal symmetry-driven contrast mechanism for atomic resolution imaging is applicable to 4D-STEM images that were recorded with fast pixelated direct electron detectors [40].

Data-driven transmission electron microscopy [83] is rapidly becoming a reality, but objective analytical techniques for symmetry quantifications may always be necessary and not be replaced by machine learning techniques. The importance of including noise in the synthetic training data of machine learning systems that predict crystal orientations and map strains on the basis of experimental 4D-STEM diffraction datasets has recently been highlighted [84].

While *"universal approximators"* [85] are theoretically possible, it remains to be seen [86] if the inclusion relation problem of crystallographic symmetry classifications can be overcome with such machines at a reasonable cost, especially when there is insufficient structural information in the training data and experimental noise is present.

Classification results for plane symmetry groups, i.e. the 17 space group types in 2D, can be amazingly accurate when enough synthetic training data is used that is more or less evenly distributed over all classes. An accuracy of higher than 99 % was, for example, reported in [87] after a deep convolutional network was trained with 600,000 synthetic 2D periodic images that were generated from random selections of images of the ImageNet (https://imagenet.stanford.edu/) dataset. The validation of this machine learning system was done on an additional 200,000 of such images.

Symmetry inclusion relationships were partly revealed in the confusion matrix of that study [87]. Given the very large training and validation sets, these authors were well aware of the limitations of their neural network and concluded *"until neural networks can learn concepts like symmetry parsimoniously, their use in materials micro-structures should be restricted to narrow bounded problems (e.g., binary classification), where subsequent physics-based validation is possible, or there is an expert-in-the-loop"* [87].

The full one-million-entry dataset of the study in [87] is available online [88]. In an earlier study [89], a neural network that was trained on this "wallpaper" dataset was used to classify synthetic atomic resolution images from crystals. In order to make the synthetic atomic resolution images resemble experimental images more closely, Gaussian blur and random atomic displacements from the ideal atomic positions were applied. That image dataset had also about a million entries.

The hypothesis tested in [89] was if transfer learning [90] would enable reasonably good classification results for the study's atomic resolution images by the network that was trained on the wallpaper images and vice versa. Symmetry transfer learning did, however, not happen to any appreciable amount and the authors concluded *"if we cross-validate the performance of a model trained on one dataset with a dataset of a different type, the accuracy is no better than a random guess"* [89]. (There was also a third set of about a million 2D periodic images that were created from random noise for which analogous hypothesis test results were obtained.)

Although in a projection into one dimension, 3D space group information is contained in pair distribution functions that were calculated from powder X-ray diffraction patterns [91]. It is highly commendable to put a machine learning system for space group classifications from that kind of data into open access [92]. That system did, however, "rediscover" the symmetry inclusion problem (and offers no solution to it beyond a potentially infinitely large training data set). Its classification results are at best a bunch of space groups that are translationengleiche subgroups and supergroups of each other [91]. That system has, therefore, learned nothing about space group symmetries that has not been known for over a century.

A similar result was reported in [93] for the space group classification by a machine learning system that was trained with calculated powder X-ray diffraction patterns to which noise and background were added in order to better approximate experimental data. Three-dimensional space group information is contained in such patterns as well, again as a projection into one dimension.

The authors of [93] reported a space group classification accuracy of around 54 % on experimental data and gave their network the option to refuse a classification if it could only be made with a high uncertainty. This permission to refuse improved the classification accuracy to 82 %, albeit at the price of leaving half of the experimental data unclassified.

What may be inferred from the results of [93] is not to expect too much from a machine classifier that delivers a variational approximation to some arbitrary function while bypassing well established analytical procedures for space group classifications. In order to save time, it may, however, often be advantageous to use a machine classifier to find a set of 3D subgroups and supergroups amongst which the correct space group of an experimental X-ray powder diffraction pattern is to be found by analytical means.

One of the conclusions of a similar study using calculated powder X-ray diffraction patterns of inorganic materials [94] was that they could not train their neural network as well as they would have liked for classifications into low symmetric crystal systems and space groups. The main reason for this was probably the above-mentioned very strong imbalance in the occupation of space group symmetries for this kind of crystals. For the higher symmetric space group symmetries, these authors obtained, on the other hand, comparable (and partially superior) classification results [94] with respect to earlier [34] studies by other authors.

The relative large error rate in their classifications into low crystallographic symmetries may again suggest that there is simply not enough symmetry information in the corresponding powder X-ray diffraction patterns when the crystal symmetry is low. If that were indeed to be so, this would not bode well for classifications of 2D data from low symmetry crystals into plane symmetry group, Laue class, and Bravais lattice type. This is because lower symmetric crystals project to lower symmetric plane symmetry groups.

A comprehensive study of the application of machine learning system to the extraction of lattice parameters from powder X-ray diffraction patterns [95] concluded in 2021 that *"the presence of multiple phases, baseline noise and peak broadening are particularly damaging"* to the prediction accuracy. *"Incorporating these experimental conditions into the training is absolutely necessary"*. Nearly a million crystallographic datasets from both the Inorganic Crystal Structure Database [96] and the Cambridge Structural Database [97] were used in that study, providing both higher symmetric inorganic crystal data and lower symmetric organic crystal data. That study identifies a *"pattern recognition"* study [98] from the year 2004 as the start of the field of applying machine learning techniques to unit cell parameter prediction from power X-ray diffraction patterns without prior crystallographic indexing.

The so-called "*data science approach*" i.e. an arsenal of contemporary machine learning system, have recently also been applied to powder neutron diffractometry [99,100]. Just as in powder X-ray diffractometry, there is 3D crystal structure information in powder neutron diffraction data that is projected into one dimension.

So far it was only attempted to predict selected crystal systems for five of the seven types of metric symmetry from simulated powder neutron diffraction training data [99] and to test the machine learning system with a few temperature depended experimental datasets from $BaTiO_3$ that spanned the material's rhombohedral/(trigonal), tetragonal, and cubic phases. The synthetic training data were calculated neutron diffraction profiles for $BaTiO_3$ over a wide range of hypothetical lattice parameters for standard experimental conditions of the employed diffractometer. Most of these training data are unphysical as, depending on the temperature, this material features in nature only a few crystal phases with distinct lattice parameters (at standard pressure) [101].

The tetragonal phase of $BaTiO_3$ features a c/a lattice vector magnitude ratio of approximately 1.01, signifying a strong cubic translational/metric pseudosymmetry. Likewise, the angle between the lattice vectors is in the rhombohedral $BaTiO_3$ phase within 0.15 degrees of 90.0° [101], resulting in a cubic metric/translational pseudo-symmetry as well. Similar lattice parameter relationships are observed for the crystal phases of other compounds that crystallize in the perovskite structural prototype. It will, therefore, be difficult to prepare a machine learning system that can reliably classify the metric symmetry of experimental diffraction patterns from the crystal phases of members of the perovskite structural prototype.

Although being unphysical, the synthetic training data set allowed for highly accurate classifications of synthetic validation data into the (higher symmetric) crystal systems cubic, tetragonal, and rhombohedral/(trigonal). For synthetic monoclinic and triclinic validation data, the prediction accuracies were approximately 90 and 95 %, respectively. Most misclassifications of synthetic validation data occurred for the monoclinic crystal system. A completely unrestricted metric symmetry, i.e. that of the triclinic crystal system, was predicted for this system for about 10 % of the synthetic test patterns [100]. This demonstrates that the machine learning system worked largely as intended at the mathematical level for metric symmetry classifications of synthetic data. Due to using a largely unphysical training dataset that is based on hypothetical arrangements of the atoms of the $BaTiO_3$ compound, its usefulness is extremely limited with respect to classifications of experimental powder neutron diffraction data that were recorded with other diffractometers and from materials outside of the perovskite structural prototype.

For an experimental powder neutron diffraction dataset from cubic $BaTiO_3$, i.e. the ferroelectric phase that is stable above approximately 403 °K, the machine learning system in [99] predicted a lattice constant of 4.215 Å. The widely accepted lattice constant for this particular crystal phase is, however, 3.996 Å [101]. As there was visually a good match to the experimental powder neutron diffraction profile in [99] for the cubic crystal phase, the discrepancy might have been due to some uncorrected systematic errors in the experimental data. The rhombohedral lattice angle was in [99] predicted to be 41.6°, whereas the accepted literature value is 89.85° [101]. The predicted c/a ratio for the tetragonal lattice of $BaTiO_3$ was 1.045, for a literature value of 1.011 [101]. These results show that predictions for data that are of a different type than the training data, i.e. so called generalizations, are to be taken with more than one grain of salt.

The authors of that study were, nevertheless, very optimistic that an expert-in-the-loop [87] would in the future no longer be needed for crystallographic analyses [99], see Fig. A-4. All structural parameters, i.e. fractional atomic coordinates in the unit cell, bond length and angles, (and presumably also isotropic or anisotropic vibration parameters of the atoms), should according to these authors be predicable with a machine learning based workflow in the future.

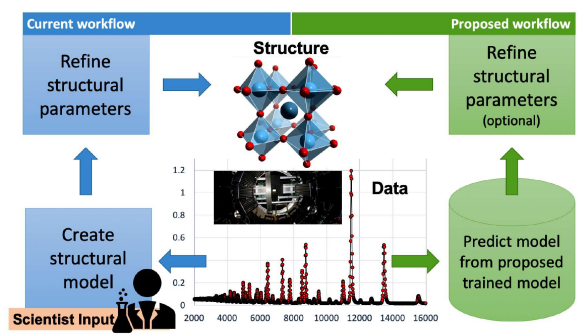

**Fig. A-4:** From [99] with permission. *Left:* Classical workflow of solving and refining a crystal structure with an expert-in-the-loop [87] who comes up with an initial model of the crystal structures and decides on the refinement strategy to make the most out of the structural information in the experimental data. This is a two-step procedure where the refinement calculations are based on the employed space group symmetry. *Middle:* from the bottom to the top, experimental time of flight neutron diffraction data, photo of a diffractometer, and cartoon of a crystal structure of the perovskite structural prototype. *Right:* Proposed one-step procedure for the determination of a crystal structure by a future machine-learning system where the structure refinement is optional.

This might perhaps work one day for very simple crystal structures that are mono-atomic, since each atom features three positional parameters and up to six vibrational parameters per unit cell. Note that the proposed "workflow of the future" in Fig. A-4 does not even involve a mandatory crystal structure refinement step so that its results would probably not be very precise (not to mention accurate in the physical sense, see Fig. A-5).

This author strongly suspects that there may not be enough information in experimental powder neutron diffraction data to succeed with a complete crystal structure determination of complex chemical compounds. There seem, at present, to be no plans to build machine learning systems for space group predictions from powder neutron diffraction data [99,100], so that the determined crystal structures would presumably all be presented in space group *P1* (at least initially until an on-line tool such as FINDSYM [102] is used by an expert-in-the-loop to find a reasonably well fitting space group for the predicted atomic coordinates).

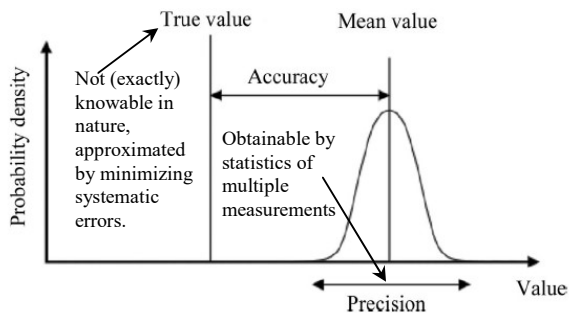

**Fig. A-5:** The concepts of precision and accuracy in experimental measurement science. From [38] with permission and annotations.

The 2020 follow up study on [99] used only synthetic training data for the trigonal, tetragonal, and cubic phases of $BaTiO_3$ [100]. The predicted lattice parameters of experimental test data were closer to the accepted literature values (of [101]), perhaps as result of a better physics-driven pre-processing of the experimental data or the restriction of the synthetic training data set to only three metric symmetries (crystal systems) of synthetic data [100]. (The predicted rhombohedral angle, was, however, still unreasonable.)

While it is true that *"existing methods ... rely on the fidelity of physics-driven forward models for accuracy"* and that the proposed machine learning alternative in [100] *"is fast, data-driven and less reliant on the fidelity of the underlying physics",* this author does not believe that *"data-driven machine learning"* is *"potentially more accurate"* [99] than standard crystallographic analyses that involve structure refinements under the constraints of an objectively determined space group. Support for their assertion that the data-driven approach is potentially more accurate than the physics-driven approach was lacking in [99,100] so that one may consider it at present as wishful thinking (rather than a meaningful contribution to the experimental science of 3D crystallography). There might be some confusion on what exactly is accuracy in machine learning studies, on the one hand, and measurement based physical science on the other hand.

Accuracy in the physics sense, see Fig. A-4, depends crucially on the skillful modeling of an experiment by the application of the pertinent physical laws [38], which the data science approach largely aims to ignore. A good test for the successful removal of systematic errors in experiments, i.e. increased accuracy, is the better agreement of results that were obtained in a statistically sound manner (with high precision) by different methods (based on different physical laws) [38].

Instead, accuracy is in machine learning often understood as simple adherence to some "ground truth", which is assumed to be known with unlimited accuracy and precision (just like a Dirac delta function in a plot of the probability density over a value such as Fig. A-4). Note in passing that the correct space group symmetry is a very powerful constraint that allows for spatial averaging over the asymmetric 3D unit, which can be up to 192 times smaller than the translation periodic unit cell.

Shannon's information theory was used more than 60 years ago to show that complete crystal structure determinations are theoretically possible based on a comprehensive set of 3D diffraction data alone when *analytic* routes are followed [30]. This author is not aware that something comparable has been demonstrated with respect to the utility of machine learning systems for crystallographic symmetry classifications and crystal structure determination.

Note that compared to a whole crystal structure determination, an accompanying space group classification is much less involved, but a different kind of problem. As mentioned above, the correct space group classification is the precondition for the objective, experimental data supported, refinement of the crystal structure.

It is well known that pseudosymmetries are not rare in real-world crystals [7-9,71-73]. Two-dimensional images with pseudosymmetries are, however, so far absent from labeled training datasets in machine learning studies [54-56,58-64], so that distinctions between genuine symmetries and pseudosymmetries cannot be achieved by such machines. Also, several of these studies did not even include noisy and experimental images into their training data sets.

One may wonder if preparing convolutional correlation-detection machines for classifications that necessarily encounter symmetry inclusion relationships were worthwhile endeavors when it may eventually turn out that this is simply outside of the capabilities of "translation-only" equivariant machine learning systems.

Geometric deep learning, as recently reviewed in [103], benefits from hardware implementations of geometric priors. Such priors capture information on the symmetry properties of the input data. Symmetry inclusion relations were, however, not specifically addresses in an equivariant network [104], where point symmetries *4* and *4mm* were combined with the inherent translation symmetry of convolutional neural networks.

A practical advantage of Euclidian equivariant neural networks is that they do not rely on input data that needed to be augmented with a multitude of arbitrarily rotated versions of the same input images [105a]. Another advantage of these machine learning systems is that they cannot make predictions that are nonsensical from a symmetry viewpoint [105b]. There are a few more machine learning papers [106a-d] on symmetry beyond pure translation equivalent neural networks. Improvements of machine learning systems by means of the incorporation of information theoretic concepts have also been proposed [107a,b].

It remains to be seen if "crystallographic symmetry savvy" machine learning systems can be build. A symmetry savvy system should be much more sophisticated than a "symmetry aware" [87] system, as it would need to deal with the crystallographic pseudosymmetries, symmetry inclusion relationships, and experimental noise all at the same time. The question if *"artificial intelligence*[15]*"* (AI, [108]) has *"become alchemy"* or is a viable form of experimental computer engineering has recently been discussed in the journal *Science* [109]. Despite the somewhat negative connotation of the word "alchemy" in this direct quote, one needs to remember that the efforts of alchemist contributed to mankind's material basis as they did create both gun powder and porcelain in their heydays. For technological progress to be continued in the long run, a strong theoretical basis needed to be developed to underpin experimental engineering breakthroughs.

---

[15] Artificial intelligence has been defined as *"a system's ability to interpret external data correctly, to learn from such data, and to use those learnings to achieve specific goals and tasks through flexible adaptation"* [108]. It is fair to state that the above mentioned contemporary machine learning systems have "learned" something about their subjectively selected training data sets, but not their essence, i.e. the crystallographic symmetries themselves into which noisy experimental data is to be classified in the presence of pseudosymmetries, experimental noise, and symmetry inclusion relationships.

Fittingly to the alchemy metaphor, a replication crisis has also been noted for the AI field as a whole [110] and progress in some sub-fields seems to have stalled [111]. For the sake of progress, one can only hope that there will not be another "*AI winter*" [112]. A first attempt to briefly review the successes and failures of the machine learning approaches to crystallographic symmetry classifications was provided by this author in the expanded on-line version of [34] in the year 2019. The discussion above may serve as some kind of an update to that review.

As the main part of this paper demonstrates, statistically sound analytic alternatives to both the computer codes of the computational symmetry community and the application of machine learning systems to the task of crystallographic symmetry classifications do exist [7,10,24,25,29]. These alternatives even outperform human experts with respect to their accuracy in the admittedly difficult distinction between genuine symmetries and strong Fedorov type pseudosymmetries in the presence of noise (as demonstrated in this paper).

As to the merits of *analytical* approaches in general, crystallographers have over the years found multiple workarounds for the recovery of the lost structure factor phases of crystal diffraction experiments (but only recently for the symmetry inclusion problem in 2D [7,10,24,25,29]). One such lost structure factor phases workaround is used in electron crystallography, where good estimates of these phases are recovered from sufficiently well resolved images of a crystal (Nobel Prizes to Sir Aaron Klug, 1982, Richard Henderson, Joachim Frank, and Jacques Dubochet, 2017).

These estimates are then used for crystallographic symmetry classifications of 2D projections of space group symmetries and the subsequent crystallographic processing of the images [113], which facilitated the solving of an inorganic crystal structure to atomic resolution by electron microscopy [114] as early as the year 1984. The projected space group was known from X-ray crystallography in [114] so that it did not need to be determined from the input data for the crystallographic processing of the experimental transmission electron microscope images of that study.

The author of [30] concluded more than six decades ago that *"information theory ... points the way to extending the resolution of detail in electron micrographs of crystals"*. In the modern terminology of "computational imaging" [20], this kind of "resolution of detail" is referred to as intrinsic image quality and structural resolution, as discussed in appendix C.

Nearly four decades ago, justifications for the validity of the direct methods approach (Nobel Prizes to Herbert A. Hauptman and Jerome Karle, 1985) of diffraction based crystallography have been derived on the basis of Shannon's information theory [115].

Around the same time, the crystallographic inversion problem, i.e. the derivation of crystal structures form the intensities of kinematic X-ray diffraction patterns from single crystals, has been formulated as a statistical inference problem [116-118]. That problem is to be solved by an optimal utilization of available partial structural information while being maximally non-committal to missing information. The term *"statistical geometry"* has been coined in that series of papers on a maximum entropy approach to the solving and refining of 3D crystal structures [116-118] that is also applicable to more or less 2D periodic images.

Applications of the maximum entropy principle in crystallography [119] and in the processing of noisy and sparsely sampled non-periodic linear images [120] have also been discussed and demonstrated at around the same time. The world had, however, to wait until 1996 for Kanatani's geometric form of information theory [26-28] and his solution to the symmetry inclusion problem to emerge.

The application of Kanatani's statistical theory to crystallographic symmetry classifications is by now reasonably well developed so that it can be taken up by the wider scientific community. In other words, there is no good justification to stick to a *"procedure"* that has in the year 2020 been ruled to be *"not sufficiently standardized"* and where *"a number of different variables (e.g. ... threshold value for interpretation) can substantially impact the outcome"* [11].

There is also no good justification to wait for robots to do accurate crystallographic symmetry classifications and whole crystal structure determinations for us. First off, there is probably always a human expert that needs to be in the loop [87] for the foreseeable future to make *physically accurate* crystallographic symmetry classifications and subsequent crystal structure analyses reliably. Secondly, such loops should not be closed to human beings prematurely and future AI systems need to work for the greater good rather than only maximizing profits [121].

A statistical extension of classical (space group based) crystallography has recently been proposed on the basis of Shannon's information theory [122]. In the humble opinion of this author, the future seems to belong to the *analytical* approaches. The most objective (unbiased) description of a noisy system that is not completely amenable to experimental verification is, after all at any one time, the one with maximum entropy (which minimizes the estimate of the expected Kullback-Leibler divergence) with respect to whatever has already been well established by science and mathematics.

Nevertheless, Thomas Proffen was probably right when he answered his own rhetorical question *"So will machine learning replace crystallographers?"* with *"No, but crystallographers using machine-learning techniques will likely surpass those who do not."* [123].

Computer scientists need to head the advice that *"the design of model and loss function should reflect the basic physical and chemical principles"* and that *"it is necessary to generate meaningful ground truth for network training"* [124]. Only then can studies that are nearly meaningless to crystallographers, e.g. [56], be avoided.

*Appendix G: Outlook on future developments of the information theoretic crystallographic symmetry classification and quantification methodology and their potential applications*

The assumption had to be made in the main part of this paper that there is indeed more than translation symmetry in a more or less 2D periodic image. This may, however, not always be the case. There are certainly (approximately) 2D periodic patterns with and without noise in which all point/site symmetries higher than the identity operation are only pseudosymmetries and not genuine. These patterns are revealed by large ratios of the complex Fourier coefficient residuals for all plane symmetry groups with $k_l = 2$ or 3 and large Fourier coefficient amplitude residuals for all projected Laue classes with $k_l = 4$ or 6. Those patterns would be misclassified by the author's methods at the present stage of their development if the facts were ignored that the residuals for all of these groups and classes are rather large.

Formulations of geometric information criteria are possible where the generalized noise does not need to be approximately Gaussian distributed. The sums of squared residuals in Kanatani's formulation are then to be replaced by maximal likelihood estimates that are specific to that noise distribution. The geometric-model selection-bias correction terms need to be specific to the distribution of that noise also.

Crystallographic studies of the quaternary structure of intrinsic membrane proteins in lipid bilayers are in the structural biology field based on parallel-illumination transmission electron microscopy (TEM) images that are dominated by Poisson distributed shot noise. As mentioned in the previous sub-section, an information theoretic approach to the classification and quantification of crystallographic symmetries in such highly beam-sensitive samples (and the digital images that were recorded from them) could be specifically developed by a generalization of Kanatani's geometric framework.

For the time being, this author sees no harm in using the method of this paper in that particular field as well. This is for two reasons: (*i*) because with moderate electron doses aggregated shot noise becomes approximately Gaussian distributed and (*ii*) the inferior/subjective traditional crystallographic symmetry classification methods (that do not model the noise at all) are currently used for exactly this purpose.

So far unpublished results of this author on the plane symmetry group and Laue class classification of the cyclic nucleotide-modulated potassium channel MloK1 from bacterium mesorhizobium loti in both the open and closed conformations indicate that the projected genuine, i.e. least broken, quaternary symmetry of this protein complex is point group 2. There is, however, a strong four-fold pseudo-symmetry along the channel axis as indicated by the relatively low squared residual of the complex Fourier coefficients for plane symmetry group *p4gm*.

This makes the potassium channel a dimer of two dimers, while other authors [125-127] claimed it to be a tetramer. Their claim relies, however, on the traditional crystallographic symmetry classification methodology, which is inherently subjective.

Incidentally, the experimental facts of this author's study on the above mentioned MloK1 potassium channel are similar to the results of the information-theoretic analysis of the noisiest crystallographic pattern in the main part of this paper. The histograms of the experimental TEM images looked rather similar to the histogram inset in Figure 6 in the sense that there was only one broad peak with a mean value that corresponded to approximately 50 % of the whole dynamic intensity range. In other words, there was enough shot noise in the experimental images that the generalized noise became approximately Gaussian distributed.

According to other authors [125-127], the projected plane symmetry of MloK1 potassium channel crystals from this bacterium in lipid bilayers is plane symmetry group *p4gm*. This author's analysis indicates, on the other hand, that this can only be a strong pseudosymmetry because projected Laue class *2mm* has been identified as the K-L best representation of the symmetry information in the amplitude maps of the discrete Fourier transforms of the TEM images. Note that this analysis was based on some[16] of the same experimental images that [126] used in their study, as downloaded from the EMDataResource [128].

A tomographic images and derived electron density maps supported model mechanism for the opening and closing of this particular potassium channel that is restricted to four-fold rotation symmetry, as the one proposed in [128] has (at the present time), accordingly, less experimental support than an alternative mechanism that is restricted to two-fold rotation symmetry only. Note that the identification of projected Laue class *2mm* as the K-L best model of the experimental data rules the existence of genuine four-fold rotation points as site symmetries in the unit cell of the MloK1 potassium channel crystal out in an analogous manner, as the entries for projected Laue class *4* in Tables 7 and 9 rule out plane symmetry group *p4gm* for the noisy crystallographic pattern in Figure 6.

It is noticeable that it was again the information theoretic projected Laue class determination that led to the identification of a strong Fedorov type pseudosymmetry at the site/point symmetry level. Presumably, projected Laue class determinations by the new method are less sensitive to noise than the corresponding plane symmetry group determinations. (Amplitude maps of discrete Fourier transforms of perfect crystallographic patterns are known to be translation invariant.)

Complementing information theoretic classification studies of transmission electron diffraction spot patterns from intrinsic membrane proteins under zero-tilt condition would be helpful as these patterns feature typically more spots than the number of Fourier coefficients of the corresponding TEM images. This means they contain more point/site symmetry specific information. Electron diffraction patterns from perfect plane-parallel crystals are theoretically translation invariant so that small random sample movements under the electron beam might be tolerable when projected point symmetry classifications are made on the basis of such patterns.

There is, thus, a motivation for the development of an information-theoretic projected point symmetry classification and quantifications method. There are, in addition, very interesting developments in 4D STEM [40] with fast pixelated electron detectors. A new symmetry-contrast imaging mode for has, for example, been recently demonstrated [41].

---

[16] Those experimental images were recorded with a large underfocus at a nominal zero-tilt setting of the specimen goniometer.

Future developments of that contrast mechanism into a widely accepted standard are, however, hampered by the symmetry inclusion relationships. The incorporation of a newly developed information-theoretic projected point symmetry group classification and quantifications method from electron diffraction patterns would solve this problem.

This author has taken up the challenge to develop such a method for selected area electron diffraction spot patterns, precession electron diffraction patterns, nearly parallel-illumination nano-diffraction disk patterns, and convergent beam micro-diffraction patterns with essentially non-overlapping and featureless (blank) electron diffraction disks.

An information theoretic approach to the quantification of the crystallographic point symmetries in electron diffraction spot and featureless disk Bragg diffraction patterns that contain 3D information is currently under development [129-132]. That method is highly sensitive to symmetry breakings so that genuine symmetries can be distinguished from pseudosymmetries in the presence of symmetry inclusion relations and experimental noise (just as the method that is described in the main part of this paper). That method is going to be probabilistic, generalized noise level dependent, robust with respect to small mis-orientations around low-index zone axes, will allow for evidence-based geometric model rankings, and be free of arbitrary set thresholds for interpretations. It has, therefore, the potential to contribute to the development of *objective* symmetry contrast mechanisms in atomic resolution imaging. For materials with medium to large unit cells, one could potentially quantify the site symmetries of most of the atomic columns in a projected unit cell by augmenting the general approach of [23] with geometric Akaike weights when sufficiently well resolved 4D-STEM data have been collected.

### *Appendix H: The Genesis of "Tiles with Quasi-ellipses, 1992", contributed to this paper by the artist Eva Knoll herself*

Peter Möck kindly requested that I write about the pattern that he used in this paper. It is one of many that I designed as a student, first in school and later in university, where I studied architecture. Initially, these pattens took the form of doodles, drawn on graph paper as a distraction in unchallenging classes, but they soon became a kind of puzzle: could I find a way to design forms that I could *predict* to be tessellating (tiling without gaps or overlaps, see Schattschneider, 1980).

I was certainly influenced by the images of work by M. C. Escher found in various books at home (for example, Lanners, 1973), but I was less interested in realism of shape than in the underlying structure. As I progressed with my experimentation, I narrowed down what I came to call my "vocabulary" to straight lines joining relatively near vertices of the square grid, plus quarters of circles embedded in squares of length 1, 2 or 3 (Knoll, 1997), as shown in Fig. A-6.

Over time, some of these designs involved more than one shape of tile, or tiles that incorporated a design on their surface. The example that intrigued Peter, found in Knoll (2003), is part of a specific series within the larger inventory, that goes back to a period when I experimented with designs, still incorporating the vocabulary, whose basic element was pre-determined as a 3 x 3 square, as shown in Figure A-7.

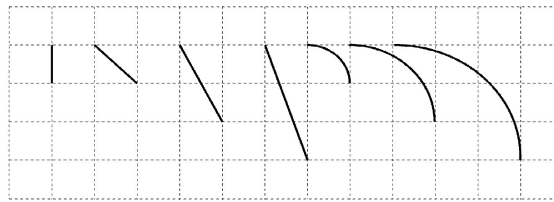

**Figure A-6:** The *Vocabulary*.

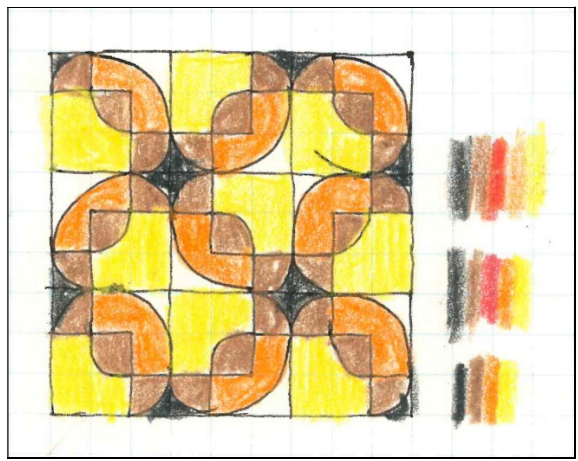

**Figure A-7:** Original sketch with colour selection.

The image that drew Peter's eye is in fact a composite of multiple copies of a digital photo of the same original tile, Figure A-8, which I had created as a bid to enter the tiling design industry.

At that time, I also had developed a preference for the *p4gm* wallpaper group and applied it often to these 3 x 3 designs, including the one selected by Peter. Figure A-6 shows the original sketch, along with some colour combinations, the last of which was used.

Multiple square-painted-ceramic tiles can not only be assembled using the square Bravais lattice type but also quasi-randomly, see Figure A-9, potentially reducing the translation symmetry of a larger assembly. This is because its specific design is entirely encapsulated in a square "fundamental region", see Figure A-8. (Because I aimed for the inclusion of a diagonal mirror line, the fundamental region of my *intended* design[17] was one half of the 3 x 3 square).

*Tiles with Quasi-Ellipses* is a bit unusual within my repertoire (www.teknollogy.com) for a few reasons. Firstly, it uses a palette of earth tones that contrasts with my more usual highly saturated choices. Second, and this is more relevant, perhaps, to Peter's work, the sharp angularity of the steps travelling diagonally creates an atypical tension when juxtaposed so closely with the curve that gave the tile its name. This curve, respecting the vocabulary, is in fact a composition of 3 quarter-circles of radius 1, 2, and 1, respectively.

---

[17] The artist's intention having been clearly stated, it is quite unreasonable to expect a perfect mirror line from a human painter. It is obvious from Figs. A-7 and A-8 that the mirror line in the original asymmetric unit is broken.

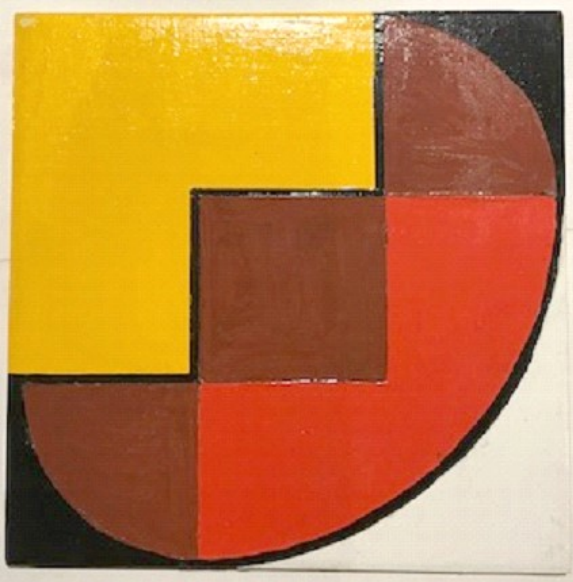

Finally, the incorporation of black outlines to mark only certain boundaries between colours signifies a departure from my practice and creates a demarcation between two multicoloured regions of the tile. The tessellation of Figure A-8 shows how this line creates interesting regions on the tiled plane, which are closed in a *p4gm* unit cell idealization but can be open in a randomised region, so that the yellow regions of adjacent tiles become joined.

### *Additional references for this appendix*

Knoll, E. *Transfert de 2-D en 3-D de l'Opus 84 de Hans Hinterreiter*. Unpublished Master's Thesis, Université de Montréal, 1997.

Knoll, E. *Life after Escher: A (Young) Artist's Journey*. In Emmer, M. and Schattschneider, D. (Eds.), Escher A Centennial Celebration, New York, Springer, 2003, 189-198.

Lanners, E. *Illusionen, Illusionen, Illusionen*. Luzern und Frankfurt/Main, Verlag C. J. Bucher, 1973.

Schattschneider, D. *Will it tile? Try the Conway Criterion*. In Mathematical Magazine, 53, 1

**Figure A-8:** Photo of the original physical tessellation element, painted with acrylic onto an unglazed, 6" (15.24 cm) edge-length tile. (The tile lays apparently on a slightly larger white piece of paper.)

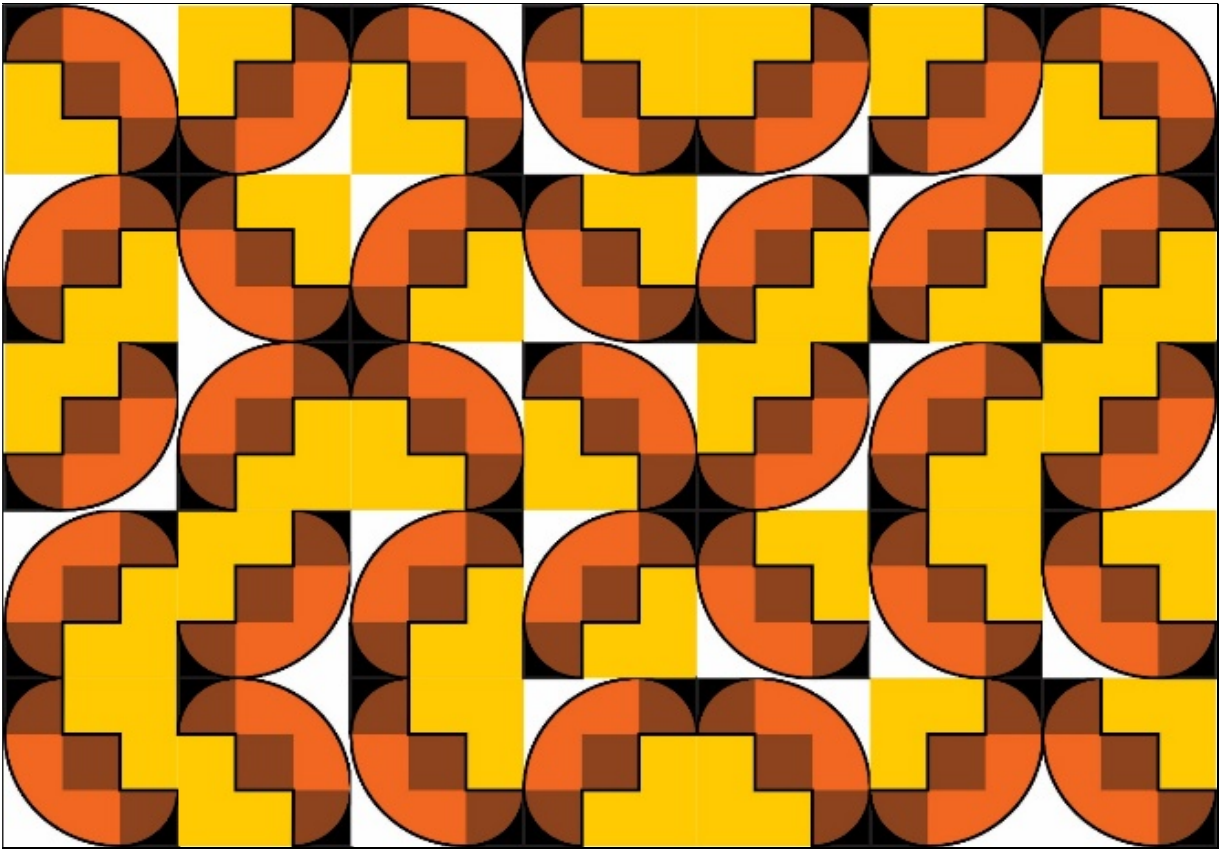

**Figure A-9:** Random assembly of the tile. Note the four tiles in the top left corner that are related by a four-fold rotation point and represent by themselves a unit cell of an ordered tessellation with plane symmetry group *p4*. Due to the size of this figure, is quite clearly revealed that the "white bow-ties" and their two parts feature only broken mirror lines.